\documentclass[aps, pra, reprint, lengthcheck, footinbib, showpacs,  citeautoscript]{revtex4-2}

\usepackage[T1]{fontenc}

\usepackage{multirow}

\usepackage{amsmath}

\usepackage{amsmath}

\usepackage{braket}
\usepackage{graphicx}

\usepackage[colorlinks, urlcolor=blue, citecolor=blue, linkcolor=blue, pdfstartview=FitH]{hyperref}
\usepackage[all]{hypcap}

\usepackage{dcolumn}
\usepackage{bm}
\usepackage{epstopdf}

\usepackage{xcolor}
\usepackage[normalem]{ulem}
\definecolor{color}{rgb}{0.11,0.45,0.02}

\newcommand{\dd}[2]{\frac{\partial #1}{\partial #2}}

\begin{document}

\title{Temperature dependence of the electron and hole Land\'e $g$-factors in CsPbI$_3$ nanocrystals in a glass matrix}

\author{Sergey~R.~Meliakov$^{1}$, Evgeny~A.~Zhukov$^{2,1}$, Vasilii~V.~Belykh$^{2}$, Mikhail~O.~Nestoklon$^{2}$,  Elena~V.~Kolobkova$^{3,4}$, Maria~S.~Kuznetsova$^5$, Manfred~Bayer$^{2}$, Dmitri~R.~Yakovlev$^{2,1}$}

\affiliation{$^{1}$P.N. Lebedev Physical Institute of the Russian Academy of Sciences, 119991 Moscow, Russia}
\affiliation{$^{2}$Experimentelle Physik 2, Technische Universit\"at Dortmund, 44227 Dortmund, Germany}
\affiliation{$^{3}$ITMO University, 199034 St. Petersburg, Russia}
\affiliation{$^{4}$St. Petersburg State Institute of Technology, 190013 St. Petersburg, Russia}
\affiliation{$^{5}$Spin Optics Laboratory, St. Petersburg State University, 198504 St. Petersburg, Russia}

\date{\today}

\begin{abstract}
The coherent spin dynamics of electrons and holes in CsPbI$_3$ perovskite nanocrystals in a glass matrix are studied by the time-resolved Faraday ellipticity technique in magnetic fields up to 430 mT across a temperature range from 6~K up to 120~K. The Land\'e $g$-factors and spin dephasing times are evaluated from the observed Larmor precession of electron and hole spins. The nanocrystal size in the three studied samples varies from about $8$ to $16$~nm, resulting in exciton transition varying from $1.69$ to $1.78$~eV at the temperature of 6~K, allowing us to study the corresponding energy dependence of the $g$-factors.  The electron $g$-factor decreases with increasing confinement energy in the NCs as result of NC size reduction, and also with increasing temperature. The hole $g$-factor shows the opposite trend. A model analysis shows that the variation of $g$-factors with NC size arises from the transition energy dependence of the $g$-factors, which becomes strongly renormalized by temperature. 
\end{abstract}

\maketitle

\textbf{Keywords:}  Perovskite nanocrystals, CsPbI$_3$, coherent spin dynamics, electron and hole $g$-factors, time-resolved Faraday ellipticity.


\section{Introduction}

The successful synthesis of colloidal nanocrystals (NCs) from the lead halide perovskites has greatly increased the possibilities for tailoring their material properties~\cite{Kovalenko2017,Efros2021}, enhancing their attractiveness for applications in photovoltaics, optoelectronics, electronics, and beyond~\cite{Shamsi2019,Dey2021,Mu2023,Huang2023}. Lead halide perovskite NCs are commonly synthesized by colloidal chemistry in solutions and can be composed of hybrid organic-inorganic or fully-inorganic materials. In comparison to the hybrid organic-inorganic perovskite NCs, the fully-inorganic NCs made, e.g., of CsPbI$_3$, CsPbBr$_3$, or CsPbCl$_3$, show considerably higher stability in ambient conditions. Further enhanced stability is gained when the NCs are synthesized in a glass matrix from a melt~\cite{Li2017,Liu2018,Liu2018a,Ye2019,Kolobkova2021,Belykh2022}, which facilitates also their usage in optoelectronic applications.

Lead halide perovskite NCs demonstrate also interesting spin-dependent properties~\cite{Vardeny2022_book}, which have been addressed by several optical and magneto-optical techniques. Among them are optical orientation and optical alignment~\cite{Nestoklon2018}, polarized emission in magnetic field~\cite{canneson2017}, time-resolved Faraday/Kerr rotation~\cite{Crane2020,Grigoryev2021,Lin2022,Meliakov2023NCs,kirstein2023_SML}, time-resolved differential transmission~\cite{Strohmair2020},  and optically-detected nuclear magnetic resonance~\cite{kirstein2023_SML}. The reported spin dynamics cover wide temporal ranges from a few picoseconds at room temperature~\cite{Strohmair2020} up to tens of nanoseconds for the spin coherence~\cite{kirstein2023_SML} and spin dephasing~\cite{Grigoryev2021} times and up to sub-milliseconds for the longitudinal spin relaxation times~\cite{Belykh2022} at cryogenic temperatures. Among the above-mentioned techniques, time-resolved Faraday/Kerr rotation is particularly informative, as it provides comprehensive data on the electron and hole Land\'e $g$-factor, spin coherence, spin dephasing and longitudinal spin relaxation. It can reveal signals from long-living electrons and holes provided by photocharging of NCs~\cite{Grigoryev2021,Meliakov2023NCs}. In CsPbBr$_3$ NCs, carrier spin coherence up to room temperature~\cite{Crane2020,Meliakov2023NCs} and its optical manipulation have been demonstrated~\cite{Lin2022}.  The spin mode-locking effect reported recently for CsPb(Cl,Br)$_3$ NCs in a glass matrix demonstrates that advanced protocols of coherent spin synchronization can be implemented in perovskite NCs~\cite{kirstein2023_SML}.

A key parameter in spin physics is the Land\'e $g$-factor, which determines the Zeeman splitting of charge carriers. We showed recently experimentally and theoretically that in  bulk lead halide perovskites the electron, hole, and exciton $g$-factors follow universal dependences on the band gap energy~\cite{kirstein2022nc,Kopteva_gX_2023}. In NCs, additional mixing of the band states by confinement gives a considerable contribution to the electron $g$-factor causing deviations from the universal dependence for bulk, but the mixing has only a weak effect on the hole $g$-factor~\cite{nestoklon2023_nl}, as predicted theoretically and confirmed experimentally in low-temperature measurements at $T=5$~K for CsPbI$_3$ NCs in glass. It is interesting to extend these experiments to higher temperatures, which could provide additional information on the band structure influence. At first glance, one would expect that the $g$-factor behaves according to the temperature shift of the band gap energy. However, it was shown for GaAs and CdTe semiconductors that the temperature dependence of the electron $g$-factor may have also other strong contributions~\cite{oestreich1995,oestreich1996,Zawadzki2008,Hubner2009}, which origin is not yet fully clarified even for conventional semiconductors.

In this paper, we study the coherent spin dynamics of electrons and holes in perovskite CsPbI$_3$ NCs in glass by time-resolved Faraday ellipticity. The spin dynamics are measured in the temperature range of $6-120$~K, from which the electron and hole $g$-factors as well as the spin relaxations times are evaluated. By analyzing NCs of different sizes in three samples at different temperatures, we measure the spin properties across a range of optical transition energies exceeding 100~meV. Our model analysis explains the observed qualitative trends for the carrier $g$-factor dispersion and its variation with temperature. Experiment and theory are also in good quantitative agreement for the hole $g$-factor, but show significant deviations for the electron $g$-factor, which origins need to be understood further.

\section{Experimental results}

We study experimentally a set of CsPbI$_3$ NCs embedded in a fluorophosphate glass matrix, namely, three samples with different NC sizes covering the range of $8-16$~nm. We label them in this paper as the samples \#1, \#2 and \#3. The photoluminescence and absorption spectra of these samples at the temperature of $T=6$~K are presented in the ESI Figure~\ref{figSI:PL-Abs}. The rather wide size dispersion in each sample together with the change between the samples allows us to cover the spectral range of exciton transitions of $1.69-1.78$~eV at 6~K (Figure~\ref{fig:6K}d), see also Ref.~\onlinecite{nestoklon2023_nl}, where samples from the same synthesis procedure were investigated. By tuning the laser photon energy, we selectively address NCs with a specific exciton transition energy corresponding to a specific NC size. For that, spectrally narrow laser pulses with 1~meV width and 1.5~ps duration are used for excitation. 

The time-resolved Faraday ellipticity (TRFE) technique is used to measure the coherent dynamics of electron and hole spins in a magnetic field to determine their parameters. This all-optical pump-probe technique exploits polarized laser pulses~\cite{Yakovlev_Ch6,belykh2019}, where spin-oriented carriers are photogenerated by the circularly-polarized pump pulses and the dynamics of their spin polarization are detected through the change of the ellipticity of the linearly polarized probe pulses~\cite{YugovaPRB09,Glazov2010}. The presented experiments are performed at cryogenic temperatures in the range of $6-120$~K. Magnetic fields up to 430~mT are applied in the Voigt geometry, perpendicular to the light wave vector.

\subsection{Electron and hole spin dynamics at $T=6$~K}

\begin{figure*}[hbt!]
\centering
\includegraphics[width=2\columnwidth]{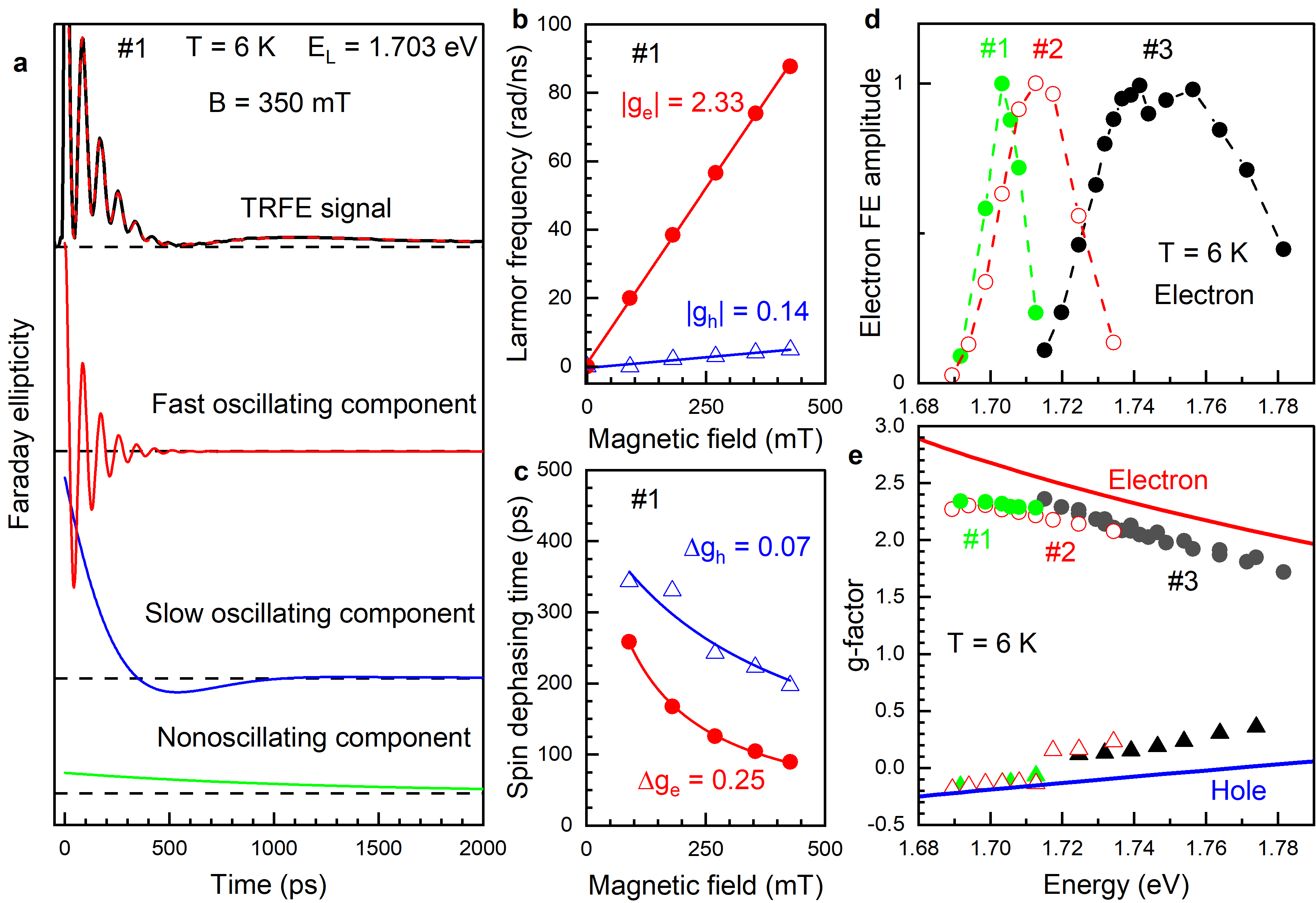}
\caption{Coherent spin dynamics in CsPbI$_3$ NCs at $T=6$~K.
(a) TRFE dynamics (black solid line) of sample \#1, measured in $B=350$~mT at the laser photon energy $E_\text{L}=1.703$~eV. Red dashed line shows a fit to the signal, using eq.~\eqref{eq:Voigt}. The three lower traces show the decomposed contributions to the signal.
(b) Magnetic field dependences of the electron (red circles) and hole (blue triangles) Larmor precession frequencies. Fits to the experimental data with eq.~\eqref{eq:LarmFreq} (solid lines) gives $|g_{\rm e}| = 2.33$ and $|g_{\rm h}| = 0.14$.
(c) Magnetic field dependences of the spin dephasing times $T_2^*$ for electrons (red circles) and holes (blue triangles). Solid lines show fits with eq.~\eqref{eq:InhDeph}.
(d) Spectral dependences of the electron FE amplitude $S_{\rm 0,e}$ for the studied samples \#1 (green circles), \#2 (red open circles), and \#3 (black circles) in $B=430$~mT. Lines are guides to the eye.
(e) Spectral dependences of the electron (circles) and hole (triangles) $g$-factors in the three studied samples. Solid lines give the calculations for CsPbI$_3$ NCs from Ref.~\onlinecite{nestoklon2023_nl}.
}
\label{fig:6K}
\end{figure*}

An example of the TRFE dynamics measured on sample \#1 at $T=6$~K for the laser photon energy $E_{\rm L}=1.703$~eV is shown at the top of Figure~\ref{fig:6K}a. The dynamics are measured with the magnetic field $B=350$~mT applied perpendicular to the light wave vector (Voigt geometry). The TRFE signal contains two oscillating and one nonoscillating component. The oscillations are observed only in finite external magnetic field and we assign them to coherent spin precession of the charge carriers with the Larmor precession frequency $\omega_\text{L}$, which is determined by the $g$-factor and scales with the magnetic field $B$ according to:
\begin{equation}
\omega_\text{L} = |g|{{\mu}_\text{B}}B/{\hbar}.
\label{eq:LarmFreq}
\end{equation}
Here, ${\mu}_\text{B}$ is the Bohr magneton and $\hbar$ is the reduced Planck constant. The decay of the oscillations is described by the spin dephasing time $T_2^*$, which in an inhomogeneous ensemble of NCs is commonly determined by the spread of Larmor precession frequencies. The dashed red line in Figure~\ref{fig:6K}a shows a fit to the spin dynamics using the following function:
\begin{multline}
A_{\rm FE}(t) \propto \sum\limits_{i=\text{e,h}} S_\text{0,i} \cos(\omega_\text{L,i}t) \exp(-t/{T_\text{2,i}^*})  +  \\
+ S_\text{1} \exp(-t/{\tau_{\rm no}}).
\label{eq:Voigt}
\end{multline}
Here, $S_\text{0,e}$ and $S_\text{0,h}$ are the initial light-induced spin polarizations of electrons and holes, respectively. $S_1$ is the initial spin polarization corresponding to the nonoscillating component. $T_\text{2,e}^*$ and $T_\text{2,h}^*$ are the electron and hole spin dephasing times, $\tau_{\rm no}$ is the decay time of the nonoscillating component. The three contributions after decomposition are shown in Figure~\ref{fig:6K}a. One can see that the nonoscillating component with $\tau_{\rm no}=1.3$~ns has a very small amplitude so that we do not consider it any further in our analysis. We assign the fast and slow oscillating components to electron ($\omega_\text{L,e} = 73.3$~rad/ns, $T_\text{2,e}^* = 100$~ps) and hole ($\omega_\text{L,h} = 4.0$~rad/ns, $T_\text{2,h}^* = 200$~ps) spin precession, respectively. We base this assignment on the known dependences of the carrier $g$-factors on the band gap energy in lead halide perovskite bulk crystals~\cite{kirstein2022nc} and on the confinement energy in nanocrystals~\cite{nestoklon2023_nl}. The spin dynamics measured in different magnetic fields on samples \#1 and \#2 are shown in the ESI Figures~\ref{figSI:MagneticEK31} and \ref{figSI:EK7}.

The magnetic field dependences of the Larmor precession frequencies are shown in Figure~\ref{fig:6K}b. As expected, they show linear dependences on the magnetic field strength without any offset for extrapolation to zero field which may arise from exchange interaction effects between electron and hole. According to eq.~\eqref{eq:LarmFreq}, the slopes of these dependences correspond to the absolute values of the electron and hole $g$-factors $|g_{\rm e}| = 2.33$ and $|g_{\rm h}| = 0.14$. It was shown in  Ref.~\onlinecite{nestoklon2023_nl}, that in CsPbI$_3$ NCs the electron $g$-factor is positive, while the hole $g$-factor crosses zero in about the studied spectral range and can be either negative or positive. The TRFE technique does not provide direct information on the $g$-factor sign. It can, in principle, be identified through dynamic nuclear polarization detected via TRFE, as demonstrated for FA$_{0.9}$Cs$_{0.1}$PbI$_{2.8}$Br$_{0.2}$ crystals, but is very sensitive to temperature and was observed only at $T=1.6$~K~\cite{kirstein2022am}. We show below that the analysis of the spectral dependence of the hole $g$-factor allows us to conclude that at the energy of 1.703~eV, $g_{\rm h}$ is negative, i.e. $g_{\rm h}=-0.14$.

It is worth to highlight that the magnetic field dependences of the Larmor precession frequencies shown in Figure~\ref{fig:6K}b  have no offset at zero magnetic field. Therefore, we can safely assign the measured spin dynamics to independent resident electrons and holes confined in the NCs and not to carriers bound in excitons. More arguments along that line can be found in Refs.~\onlinecite{Grigoryev2021,nestoklon2023_nl}. The resident carriers in the NCs can appear from long-living photocharging, where either the electron or the hole from a photogenerated electron-hole pair escapes from the NC. As a result, some NCs in the ensemble are charged with electrons, some NCs with holes, while the rest stays neutral.

The magnetic field dependences of the spin dephasing times of electrons ($T_\text{2,e}^*$) and holes ($T_\text{2,h}^*$) are shown in Figure~\ref{fig:6K}c.  With magnetic field growing from 90~mT to 430~mT, both times decrease, namely, for the electrons from 260~ps to 90~ps  and for the holes from 350~ps to 200~ps. This behavior is typical for inhomogeneous spin ensembles with a finite $g$-factor spread $\Delta g$ and can be described by the following expression~\cite{Yakovlev_Ch6}:
\begin{equation}
\frac{1}{T_2^*(B)} {\approx} \frac{1}{{T_2^*}(0)} + \frac{{{\Delta}g}{{\mu}_\text{B}}B}{\hbar}.
\label{eq:InhDeph}
\end{equation}
Here, ${T_2^*}(0)$ is the spin dephasing time at zero magnetic field. Fitting the experimental data with eq.~\eqref{eq:InhDeph}, yields $\Delta g_{\text{e}}=0.25$ for the electrons and $\Delta g_{\text{h}}=0.07$ for the holes. Thus, the relative spreads are $\Delta g_{\text{e}} / g_{\text{e}}=11\%$ and $\Delta g_{\text{h}} / g_{\text{h}}=50\%$.

Figure~\ref{fig:6K}d shows the spectral dependence of the electron FE amplitude $S_\text{0,e}$ for the three studied samples. These profiles can be considered as those of exciton absorption in the inhomogeneous NC ensembles. They have different widths with the narrowest distribution in sample \#1 and the largest distribution in sample \#3. The traces of three samples overlap spectrally with each other so that they continuously cover the spectral range of $1.69-1.78$~eV at $T=6$~K, in correspondence to the variation of NC sizes from about $8$ to $16$~nm~\cite{nestoklon2023_nl}, where the smallest NCs have the highest energy due to strongest carrier quantum confinement.

The spectral dependences of the electron and hole $g$-factors are shown in Figure~\ref{fig:6K}e. Both dependences show monotonic changes with increasing energy. The electron $g$-factor decreases from $+2.34$ to $+1.72$ and the hole $g$-factor increases starting from the negative value of $-0.19$ and reaching the positive value of $+0.36$. The experimental results are in good agreement with the predictions of the tight-binding calculations from Ref.~\onlinecite{nestoklon2023_nl} shown by the solid lines, which account for the band mixing provided by electron and hole quantum confinement in CsPbI$_3$ NCs. Note, that in Ref.~\onlinecite{nestoklon2023_nl} we measured the same CsPbI$_3$ NCs, but with a laser system having a high pulse repetition rate of 76~MHz, which has about three orders of magnitude smaller peak power. This limited the spectral range where the spin parameters could be reliably determined, e.g., for the hole $g$-factor it was only $1.68-1.72$~eV compared to $1.68-1.78$~eV in the present study, which limits systematic studies in the earlier work.

As mentioned above, from the Larmor precession frequency measured in TRFE experiments one can evaluate only the magnitude of the $g$-factor, but not its sign. The question about the sign is of particular importance for the hole $g$-factor, which is expected to cross zero in the considered spectral range. In the ESI Figure~\ref{figSI:sign} we plot the dependence of $|g_{\rm h}|$ on the laser photon energy $E_\text{L}$, revealing a nonmonotonous behavior: $|g_{\rm h}|$ decreases in the spectral range from about 1.69~eV to 1.71~eV, but increases in the range from 1.71~eV to 1.78~eV. As the monotonic dependence was theoretically predicted~\cite{nestoklon2023_nl}, we suggest that $g_{\rm h}<0$ for $E_\text{L}<1.71$~eV, crosses zero at $E_\text{L} \approx 1.71$~eV, and $g_{\rm h} >0$ for $E_\text{L}>1.71$~eV. We use this reasoning for plotting the data in Figure~\ref{fig:6K}e.

\begin{figure*}[hbt!]
\centering
\includegraphics[width=2\columnwidth]{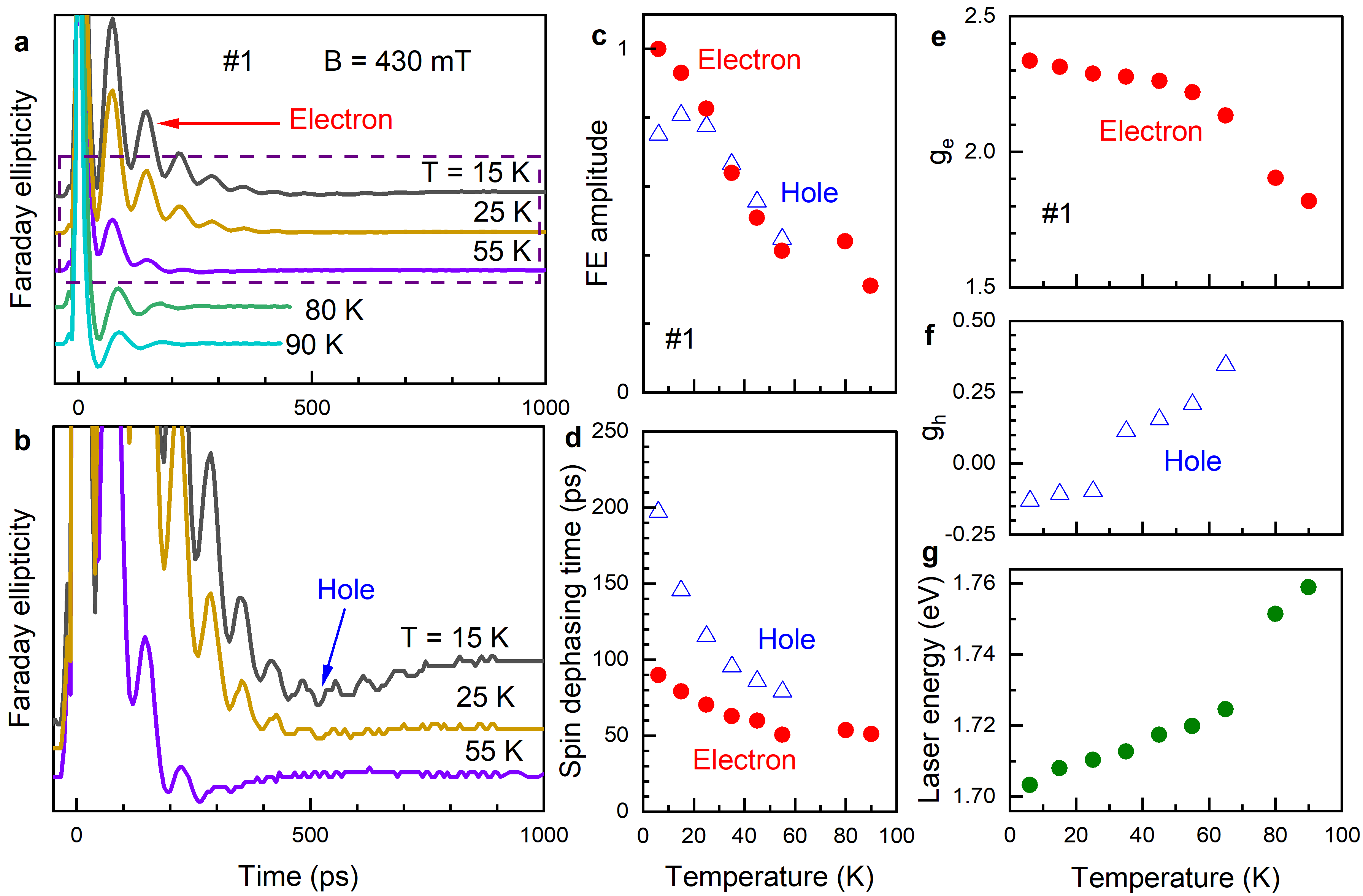}
\caption{Temperature dependence of the spin dynamics in sample \#1 for $B=430$~mT.
(a) TRFE traces at various temperatures from 15 to 90~K.
(b) Zoom into dynamics from panel (a) at 15~K, 25~K, and 55~K to highlight the hole Larmor precession.
(c) Temperature dependence of electron (red circles) and hole (blue triangles) FE amplitude. 
(d) Temperature dependence of electron and hole spin dephasing time $T_2^*$.
Temperature dependence of electron (e) and hole (f) $g$-factors and corresponding laser photon energies (g).
}
\label{fig:Temperature}
\end{figure*}

\subsection{Temperature dependence of spin dynamics}

We next turn to the spin dynamics measured at various temperatures in order to extract the temperature dependences of the spin related parameters in the CsPbI$_3$ NCs. As the temperature changes, the band gap of the material also changes and one needs to tune the laser accordingly to ensure that NCs of the same size are addressed at the different temperatures. This is difficult for strongly inhomogeneous samples, like for sample \#3. Therefore, we start with the basic temperature trends for the most homogeneous sample \#1. In what follows, we measure the complete spectral dependences at various temperatures for all samples.

Figures~\ref{fig:Temperature}a and \ref{fig:Temperature}b show the spin dynamics in sample \#1 measured at various temperatures. The laser photon energy is adjusted to the maximum of the TRFE amplitude for each temperature, as shown in Figure~\ref{fig:Temperature}g. At $T \leq 15$~K the FE dynamics consist of electron, hole and nonoscillating components. The amplitude $S_1$ of the nonoscillating component decreases with temperature, and at $T \geq 55$~K it is not detected, while the pronounced electron spin oscillations are seen up to $T=90$~K. The hole spin oscillations almost disappear around $T=25$~K, when the laser photon energy is equal to 1.71~eV. Figure~\ref{fig:6K}e shows that the hole $g$-factor crosses zero around this energy. With further temperature increase, the hole spin oscillations become prominent again and are observed up to 65~K. Note that on another spot of this sample \#1 we can measure spin oscillations up to 120~K, see the ESI Figure~\ref{figSI:temp120}.

The temperature dependences of the FE amplitudes and spin dephasing times, evaluated by fitting the spin dynamics with eq.~\eqref{eq:Voigt}, are presented in Figures~\ref{fig:Temperature}c and \ref{fig:Temperature}d, respectively. The FE amplitudes of electrons and holes are close to each other in the whole temperature range and decrease with increasing temperature. The electron spin dephasing time shortens from 90~ps at $T=6$~K to 50~ps at 90~K. The hole spin dephasing time decreases from  200~ps at $T=6$~K to 80~ps at 55~K. We suggest that the shortening of the spin dephasing times is due to acceleration of the carrier spin relaxation via their interaction with phonons at elevated temperatures.

\subsection{Temperature dependence of $g$-factors}

The temperature dependences of the electron and hole $g$-factors are shown in Figures~\ref{fig:Temperature}e and~\ref{fig:Temperature}f. They are measured at the maximum of the FE amplitude, which in sample \#1 shifts from 1.703~eV at 6~K to 1.759~eV at 90~K, see Figure~\ref{fig:Temperature}g. The electron $g$-factor decreases with temperature from +2.34 to +1.82. The hole $g$-factor is equal to $-0.14$ at $T=6$~K, crosses zero at about 30~K ($E_\text{L} \approx 1.710$~eV), and reaches $+0.34$ at 65~K ($E_\text{L}= 1.725$~eV). This  behavior is in line with the trend in the spectral dependences of the $g$-factors shown in Figure~\ref{fig:6K}e. Namely, with increasing energy, independent of whether achieved by laser energy tuning or temperature shift, the electron $g$-factor decreases and the hole $g$-factor increases.

\begin{figure}[hbt!]
\includegraphics[width=1\columnwidth]{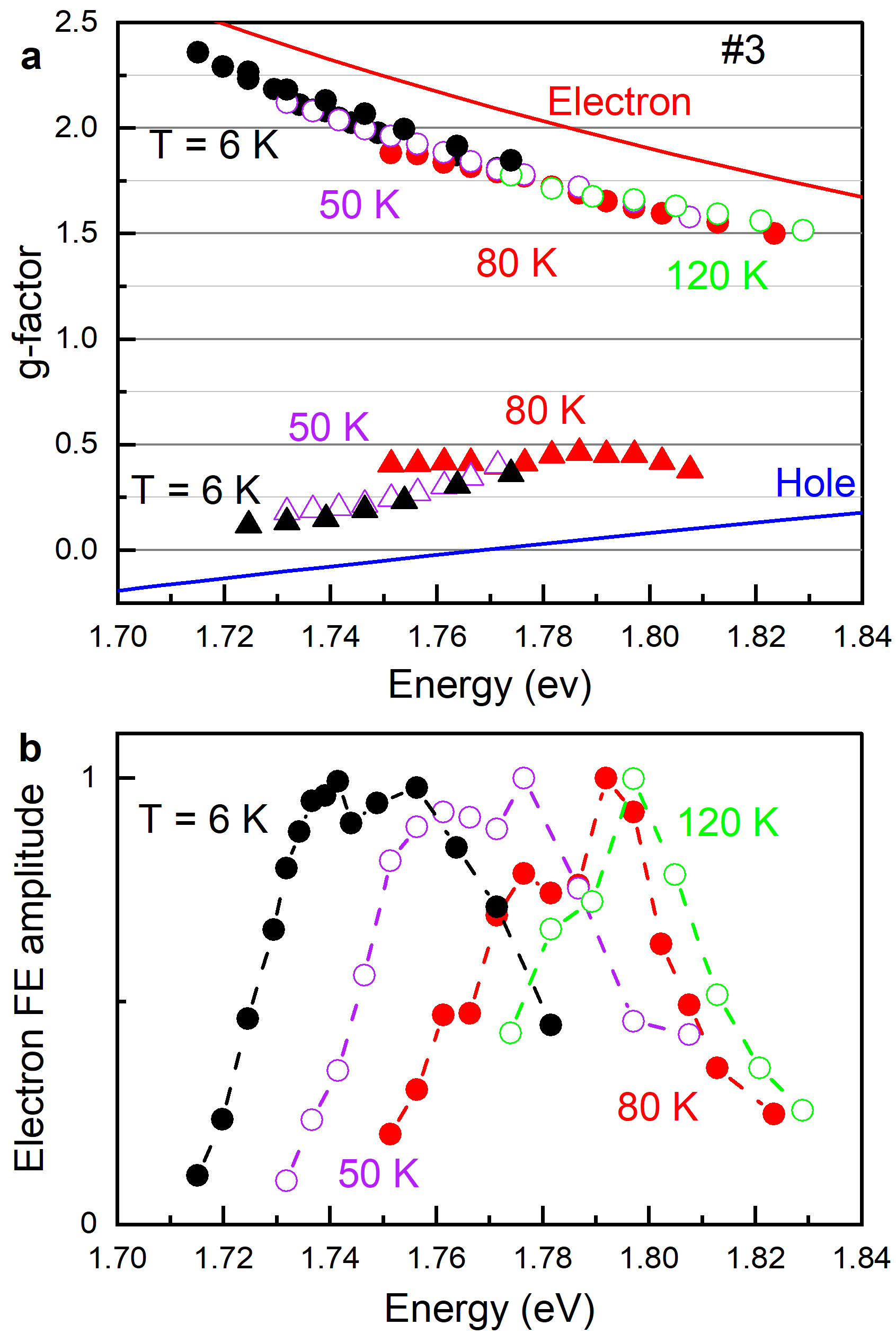}
\caption{Spectral and temperature dependences of the charge carrier $g$-factors in CsPbI$_3$ NCs of sample \#3. (a) Spectral dependences of the electron (circles) and hole (triangles) $g$-factors measured at the temperatures of 6~K (black), 50~K (purple), 80~K (red), and 120~K (green). Solid lines are calculations for $T=6$~K from Ref.~\onlinecite{nestoklon2023_nl}.
(b) Electron TRFE amplitudes at the corresponding laser photon energies and temperatures.
}
\label{fig:g_temp}
\end{figure}

\begin{figure}[hbt!]
\includegraphics[width=1\columnwidth]{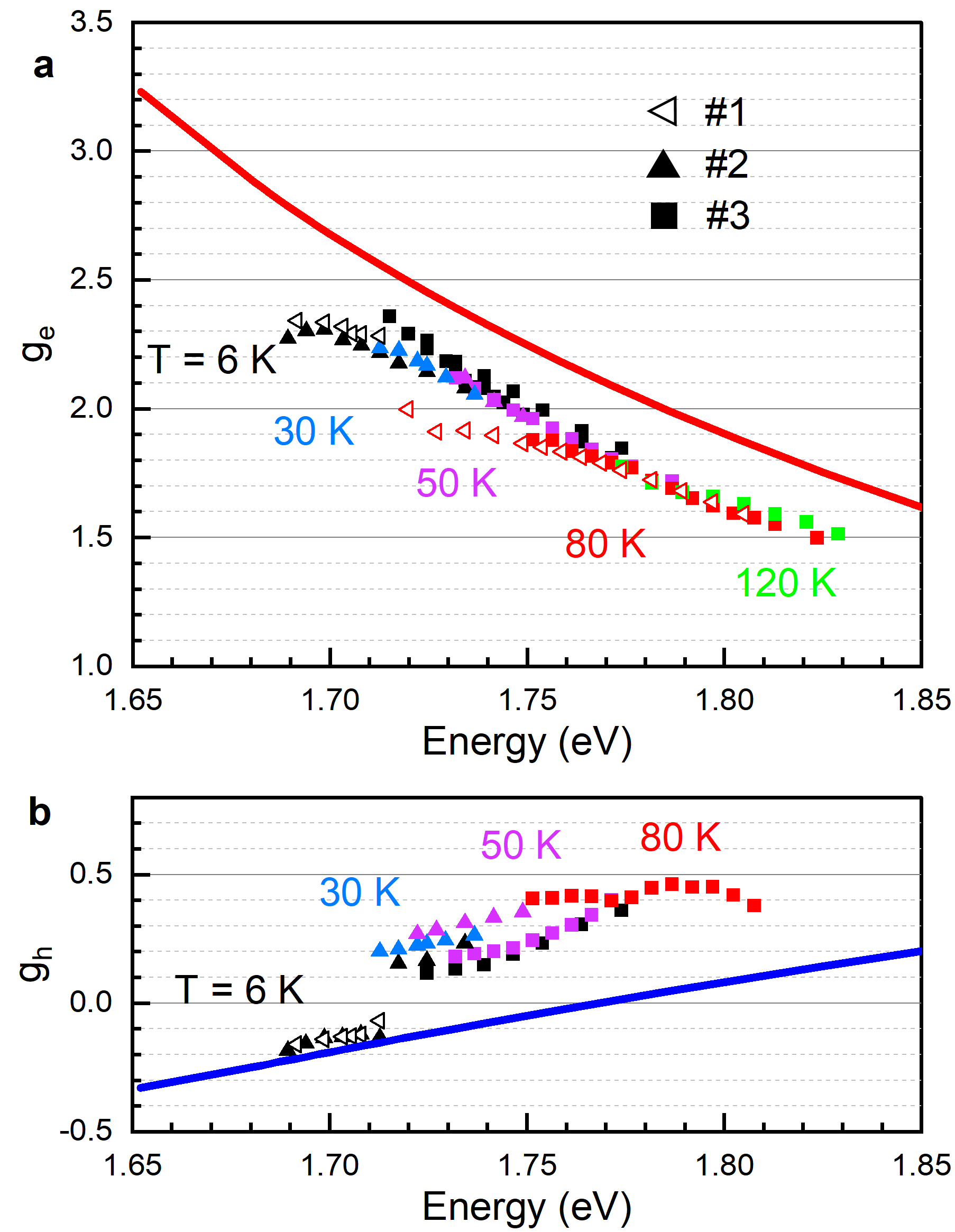}
\caption{Spectral dependences of the electron (a) and hole (b) $g$-factors in samples \#1 (open triangles), \#2 (closed triangles) and \#3 (squares) at the temperatures of 6~K (black symbols), 30~K (blue symbols), 50~K (purple symbols), 80~K (red symbols), and 120~K (green symbols). Solid lines are calculations for $T=6$~K  from Ref.~\onlinecite{nestoklon2023_nl}.
}
\label{figSI:g_comp}
\end{figure}

For closer insight into the mechanisms underlying the temperature dependence of the carrier $g$-factors we measure the spectral dependences of the $g$-factors at various temperatures. These results for sample \#3 with the largest spectral broadening are shown in Figure~\ref{fig:g_temp}a for temperatures of 6, 50, 80, and 120~K. The corresponding spectral profiles of the FE amplitude are given in Figure~\ref{fig:g_temp}b. One sees that the total covered energy range increases up to $1.68-1.83$~eV due to tuning with temperature. All gained experimental results on the carrier $g$-factors for the three studied samples and several temperatures from 6 to 120~K are collected in Figure~\ref{figSI:g_comp}. They are in agreement with the conclusions drawn for sample \#1. Namely, the electron $g$-factor decreases with increasing energy, while the hole $g$-factor increases.

\section{Modeling and discussion}

Let us analyze the factors that contribute to the temperature dependence of the electron and hole $g$-factors in CsPbI$_3$ NCs. As  starting point we take the universal dependence of carrier $g$-factors on the band gap found experimentally and confirmed theoretically for lead halide perovskite bulk crystals at $T= 5$~K \cite{kirstein2022nc}. We recently extended these studies to address the role of quantum confinement of charge carriers in NCs on the $g$-factors. The calculations predict that in NCs the hole $g$-factor follows the universal dependence, but the electron $g$-factor significantly deviates from the bulk trend~\cite{nestoklon2023_nl}. We demonstrated this expectation experimentally for CsPbI$_3$ NCs at $T=5$~K in Ref.~\onlinecite{nestoklon2023_nl} and confirmed it in the present study for a much wider spectral range, see Figure~\ref{fig:6K}e. At cryogenic temperatures, good agreement between experiment and theory is found, which is remarkable, as the band parameters for lead halide perovskite crystals and NCs are not known with high precision.

To model the temperature modifications of the $g$-factors, information on the temperature dependence of the band parameters, the band gap energy, the band dispersion determining the carrier effective masses, the band mixing at finite wave vectors for confined carriers, etc. need to be known. Also carrier-phonon polaron formation may contribute. At present, the corresponding available information is limited. Also, the understanding of the band structure details of the lead halide perovskite crystal and NCs are far from being complete. Therefore, it is not possible to perform precise calculations for the temperature dependence of the $g$-factors. Instead, we perform estimations based on the available band structure information, which details can be found in the ESI section~\ref{sec:SI:Theory}. Our goal is to explore what factors are of importance and how close the model predictions can describe the experimental results. 

We analyze the relatively small changes of the $g$-factors with energy, which arise from the temperature variations. Therefore, we consider them in the linear regime. For that the results of Ref.~\onlinecite{nestoklon2023_nl} may be compiled in the following trends:
\begin{equation}
\dd{g_e}{E_g} = \zeta_e^{b}\,,\;\;\; \dd{g_e}{E_e} =  \zeta_e^{qc}\,,\;\;\;
    \dd{g_h}{E_h} \approx \dd{g_h}{E_g} = \zeta_h\,.
\end{equation}
Here $E_g$ is the bulk band gap, $E_{e,h}$ are the energies of quantum confinement of electrons and holes.
Atomistic calculations predict that $\zeta_e^{qc} \gg \zeta_e^{b}$, which is in good agreement with experimental data. Moreover, this result is qualitatively reproduced in {\bf k}$\cdot${\bf p} calculations \cite{nestoklon2023_nl}.

\begin{figure*}[hbt!]
\includegraphics[width=1\textwidth]{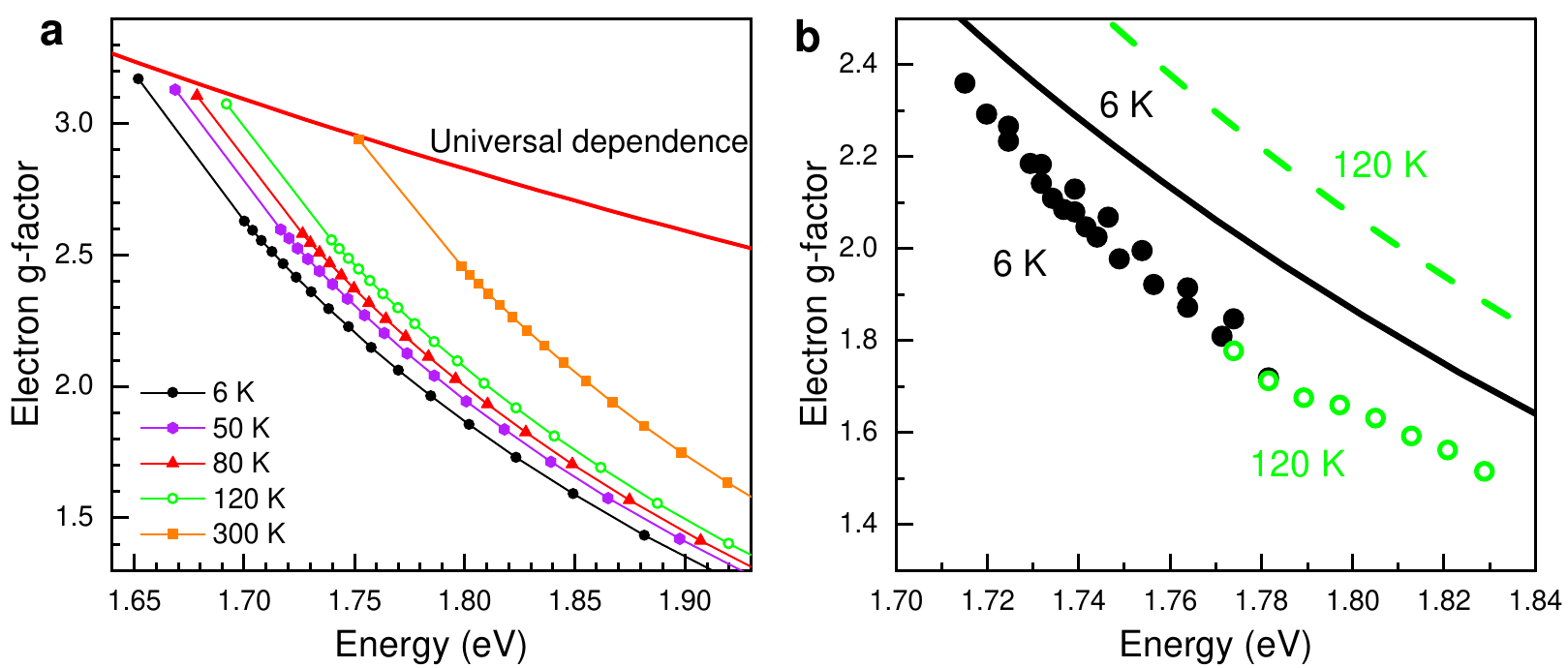}
\caption{(a) Electron $g$-factors calculated in the empirical tight-binding model following Ref.~\onlinecite{nestoklon2023_nl} for CsPbI$_3$ NCs at different temperatures, see text for details. (b) Black and green open dots show the electron $g$-factors measured for sample \#3 for the temperatures of 6~K and 120~K, respectively. Solid black and dashed green lines show the $g$-factors calculated in the empirical tight-binding model, see panel (a).   
}
\label{fig:ge_vs_T}
\end{figure*}

The temperature variation changes both the band gap and the quantum confinement energy. The bulk band gap change is linear with temperature \cite{Yang2017}. The quantum confinement energy varies due to temperature expansion of the lattice and, therefore, the NC size and also due to the temperature variation of the carrier effective mass. The temperature expansion coefficient for CsPbI$_3$ is almost independent of the crystal phase. The temperature dependence of the effective masses is not well analyzed in the literature. In a simple two-band model, the effective mass of electrons and holes should increase with increase of the bulk band gap~\cite{Yang2017}. 

The hole $g$-factor as function of energy is expected to follow the bulk trend, as we showed in Ref.~\onlinecite{nestoklon2023_nl}. Quantum confinement does not change this result qualitatively. In accordance with the theory predictions, the temperature shifts the energies to larger values and the $g$-factors slightly increase, see Figures~\ref{fig:g_temp}a and \ref{figSI:g_comp}b. This prediction is supported by the measured data. 

For electrons, however, the situation is more complicated. We calculate the expected change of the electron $g$-factor with temperature increase, following the procedure from Ref.~\onlinecite{nestoklon2023_nl} with temperature-dependent values of the lattice constant and band gap. The variation of the band gap was accounted by a change of the tight-binding parameter $E_{pc}$ fitted to reproduce the experimental value $\dd{E_g}{T}=3.1\times 10^{-4}$~eVK$^{-1}$ from Ref.~\onlinecite{Yang2017}, which closely match the temperature shifts that we measured in the studied NCs, see ESI Figure S2.   We also consider the (almost negligible) change of the lattice constant $\dd{a_0}{T}= 3.39\times 10^{-5}$~$a_0$~K$^{-1}$ \cite{Trots2008}. In Figure~\ref{fig:ge_vs_T}a we show the calculated electron $g$-factors for different temperatures. From the atomistic calculations, it follows that the temperature variation results in a band gap change, which is reflected by a shift of the $g$-factor values following the universal dependence~\cite{kirstein2022nc}. 

In Figure~\ref{fig:ge_vs_T}b we compare the calculated variation of the $g$-factor with experimental data for temperatures of 6~K and 120~K. One can see, that at $T=6$~K the calculations and experiment are much closer to each other than at 120~K.  In theory, the main effect of a temperature increase is the change of the $g$-factor by the band gap energy, which roughly follows a universal dependence, with small changes of the slope of the electron $g$-factor as function of transition energy. In contrast, the experimental data demonstrate a strong renormalization of the electron $g$-factor with temperature (larger than expected from the theory) and a much smaller slope of the electron $g$-factor as function of the transition energy at elevated temperatures, see the ESI Figure S7.

\section{Conclusions}

We have studied the spin dynamics of charge carriers in CsPbI$_3$ perovskite nanocrystals of different size by means of time-resolved Faraday ellipticity in the temperature range from 6~K up to 120~K. The spectral dependences of the electron and hole $g$-factors correspond well to model predictions accounting for the mixing of the electronic bands with increasing confinement energy, which is accompanied by a decrease of the NC size. With increasing temperature, the NC optical transition shifts to higher energies due to  increase of the band gap energy. The $g$-factors are independent of temperature in the studied range when they are measured at the same energy. An increase in temperature shifts the spectral dependences of the $g$-factors to higher energies, qualitatively following the model predictions. Namely, the electron $g$-factor decreases and the hole $g$-factor increases with growing energy as result of an increase in temperature. It is interesting and important to check further whether this trend holds also for higher temperatures up to room temperature as well as in other lead halide perovskite NCs. As follows from the comparison, the experimentally observed $g$-factors show a stronger dependence on energy. The effect of quantum confinement on the $g$-factor value is strongly renormalized by the temperature variation.  Understanding of these details and responsible mechanisms would allow one to refine band parameters for the lead halide perovskites and their NCs.

\section{Samples and Methods}

\subsection{Samples}

The studied CsPbI$_3$ nanocrystals embedded in fluorophosphate Ba(PO$_3$)$_2$-AlF$_3$ glass were synthesized by rapid cooling of a glass melt enriched with the components needed for the perovskite crystallization. Details of the method are given in Refs.~\onlinecite{Kolobkova2021,kirstein2023_SML}. The samples of fluorophosphate (FP) glass with the composition BaI$_2$-doped 35P$_2$O$_5$--35BaO--5AlF$_3$--10Ga$_2$O$_3$--10PbF$_2$--5Cs$_2$O (mol. \%) doped with BaI$_2$ were synthesized using the melt-quench technique. The glass synthesis was performed in a closed glassy carbon crucible at the temperature of $T=1050^\circ$C. About 50~g of the batch was melted in the crucible for 30~minutes, then the glass melt was cast on a glassy carbon plate and pressed to form a plate with a thickness of about 2~mm. Samples with a diameter of 5~cm were annealed at the temperature of $50^\circ$C below $T_g=400^\circ$C to remove residual stresses. The CsPbI$_3$ perovskite NCs were formed from the glass melt during the quenching. The glasses obtained in this way are doped with CsPbI$_3$ NCs. The dimensions of the NCs in the initial glass were regulated by the concentration of iodide and the rate of cooling of the melt without heat treatment above $T_g$. Three samples were investigated in this paper, which we label \#1, \#2 and \#3. Their technology codes are EK31, EK7 and EK8, respectively. They differ in the NC sizes, which is reflected by relative spectral shifts of their optical spectra. Note, that samples from the same synthesis (same codes) were investigated in Ref.~\onlinecite{nestoklon2023_nl}, even though their optical spectra slightly differ due to spatial inhomogeneity.

The change of the NC size was achieved by changing the concentration of iodine in the melt. Due to the high volatility of iodine compounds and the low viscosity of the glass-forming fluorophosphate melt at elevated temperatures, an increase in the synthesis time leads to a gradual decrease in the iodine concentration in the equilibrium melt. Thus, it is possible to completely preserve the original composition and change only the concentration of iodine due to a smooth change in the synthesis duration. Glasses with NC sizes in range $8-16$~nm were synthesized using different synthesis time. Glasses with the photoluminescence lines centered at  1.801, 1.808 and 1.809~eV (room temperature measurements) were synthesized within 40, 35, and 30 min, respectively. NC sizes are evaluated from spectral shift of exciton line in absorption spectra.


\subsection{Time-resolved Faraday ellipticity}

To study the coherent spin dynamics of carriers we use a time-resolved pump-probe technique with detection of the Faraday ellipticity (TRFE)~\cite{Yakovlev_Ch6,Glazov2010}. Spin oriented electrons and holes are generated by circularly polarized pump pulses. The used laser system (Light Conversion) generates pulses of 1.5 ps duration with a spectral width of about 1~meV at a repetition rate of 25~kHz (repetition period 40~${\mu}$s). The laser photon energy is tuned in the spectral range of $1.65-1.85$~eV in order to resonantly excite NCs of various sizes at various temperatures.
The laser beam is split into the pump and probe beams with the same photon energies. The time delay between the pump and probe pulses is controlled by a mechanical delay line. The pump beam is modulated with an electro-optical modulator between ${\sigma}^+$ and ${\sigma}^-$ circular polarization at a frequency of 26~kHz. The probe beam is linearly polarized. The Faraday ellipticity of the probe beam, which is proportional to the carrier spin polarization, is measured as function of the delay between the pump and probe pulses using a balanced photodetector connected to a lock-in amplifier synchronized with the modulator. Both pump and probe beams have power of 0.5~mW and spot sizes of about 100$~\mu$m.
For the time-resolved  measurements the samples are placed in a helium-flow optical cryostat and the temperature is varied in the range of $6-300$~K. Magnetic field up to 430~mT is applied using an electromagnet perpendicularly to the laser beam (Voigt geometry, $\textbf{B} \perp \textbf{k}$).

\section*{Conflicts of interest}
There are no conflicts of interest to declare.

\section*{Data availability}
The data supporting this article have been included as part of the Supplementary Information.

\section*{Acknowledgments}
Research performed at the P. N. Lebedev Physical Institute was financially supported by the Ministry of Science and Higher Education of the Russian Federation, Contract No. 075-15-2021-598. E.V.K. and M.S.K. acknowledge the Saint-Petersburg State University (Grant No. 122040800257-5).

%

\textbf{AUTHOR INFORMATION}

{\bf Corresponding Authors} \\
Sergey~R.~Meliakov,  Email: melyakovs@lebedev.ru   \\
Dmitri R. Yakovlev,  Email: dmitri.yakovlev@tu-dortmund.de\\

\textbf{ORCID}\\
Sergey~R.~Meliakov         0000-0003-3277-9357 \\  
Evgeny~A.~Zhukov:          0000-0003-0695-0093 \\  
Vasily~V.~Belykh:          0000-0002-0032-748X \\ 
Mikhail~O.~Nestoklon:      0000-0002-0454-342X  \\ 
Elena V. Kolobkova:        0000-0002-0134-8434 \\  
Maria S. Kuznetsova:       0000-0003-3836-1250 \\  
Manfred~Bayer:             0000-0002-0893-5949 \\ 
Dmitri R. Yakovlev:        0000-0001-7349-2745 \\  

\clearpage

\setcounter{equation}{0}
\setcounter{figure}{0}
\setcounter{table}{0}
\setcounter{section}{0}
\setcounter{subsection}{0}
\setcounter{page}{1}
\renewcommand{\theequation}{S\arabic{equation}}
\renewcommand{\thefigure}{S\arabic{figure}}
\renewcommand{\thepage}{S\arabic{page}}
\renewcommand{\thetable}{S\arabic{table}}
\renewcommand{\thesubsection}{S\arabic{subsection}}

\begin{widetext}
\begin{center}

\section*{Supplementary Information}

\textbf{Temperature dependence of the electron and hole Land\'e $g$-factors in CsPbI$_3$ nanocrystals in a glass matrix} \\

Sergey~R.~Meliakov, Evgeny~A.~Zhukov, Vasilii~V.~Belykh, Mikhail~O.~Nestoklon,  Elena~V.~Kolobkova, Maria~S.~Kuznetsova, Manfred~Bayer, Dmitri~R.~Yakovlev

\end{center}

\subsection{Photoluminescence and absorption spectra of CsPbI$_3$ NCs}

The photoluminescence (PL) spectra are excited by a continuous-wave laser operating at the wavelength of 405~nm (3.06~eV) with a power of 0.5~mW. The absorption spectra are measured using a Cary 6000i UV Vis-NIR spectrophotometer. The spectra measured at the temperature of $T=6$~K for the three studied samples are presented in Figure~\ref{figSI:PL-Abs}. The shift to higher energies among the samples results from a decrease of the NC size from sample \#1 towards sample \#3. The significant width of the PL emission lines and of the exciton absorptions peaks indicates a significant dispersion of the NC sizes in each sample. For all samples, a Stokes shift of the PL maximum from the absorption maximum is observed. 

\begin{figure}[hbt!]
\includegraphics[width=0.5\columnwidth]{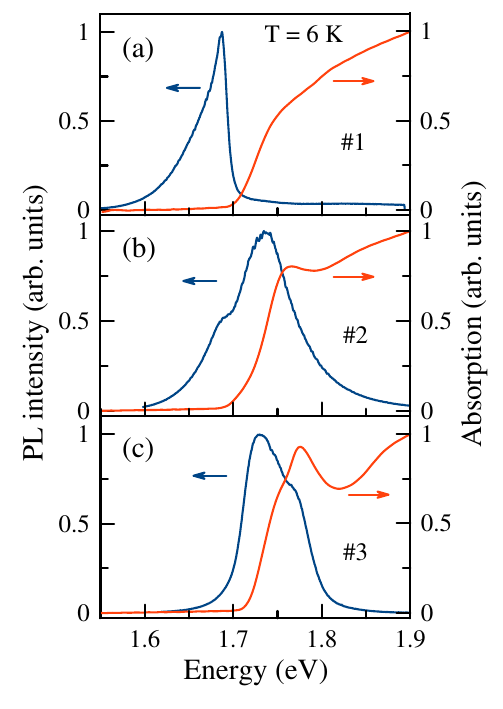}
\caption{Normalized photoluminescence (blue line) and absorption (red line) spectra of the studied CsPbI$_3$ NCs: (a) sample~\#1, (b) sample~\#2 and (c) sample~\#3. $T=6$~K.}
\label{figSI:PL-Abs}
\end{figure}

Figures ~\ref{figSI:Abs-Temp}(a,b,c) show the absorption spectra of the samples under study, measured for temperatures from 4~K up to 150~K. The edge of the absorption for all samples shifts to higher energies with increasing temperature. For samples~\#2 and \#3 the dependence of the shift on temperature, presented in Figures~\ref{figSI:Abs-Temp}(e,f), was evaluated from the shift of the maximum of the exciton absorption peak. For sample~\#1, the exciton peak is significantly broadened so that its maximum is almost impossible to determine precisely. Therefore, the temperature shift of the absorption edge was evaluated from the position of the half-maximum in the absorption edge. It should also be noted that for all samples a significant shift was observed only at temperatures above $T\simeq12$~K. At high $T$, the shift increases almost linearly with temperature. This is demonstrated by the linear fits of the dependences  in Figures~\ref{figSI:Abs-Temp}(d,e,f). The slopes of these lines are: $3.2\times 10^{-4}$~eVK$^{-1}$ (\#1), $3.8\times 10^{-4}$~eVK$^{-1}$ (\#2), and $3.5\times 10^{-4}$~eVK$^{-1}$ (\#3).

\begin{figure}[hbt!]
\includegraphics[width=0.8\columnwidth]{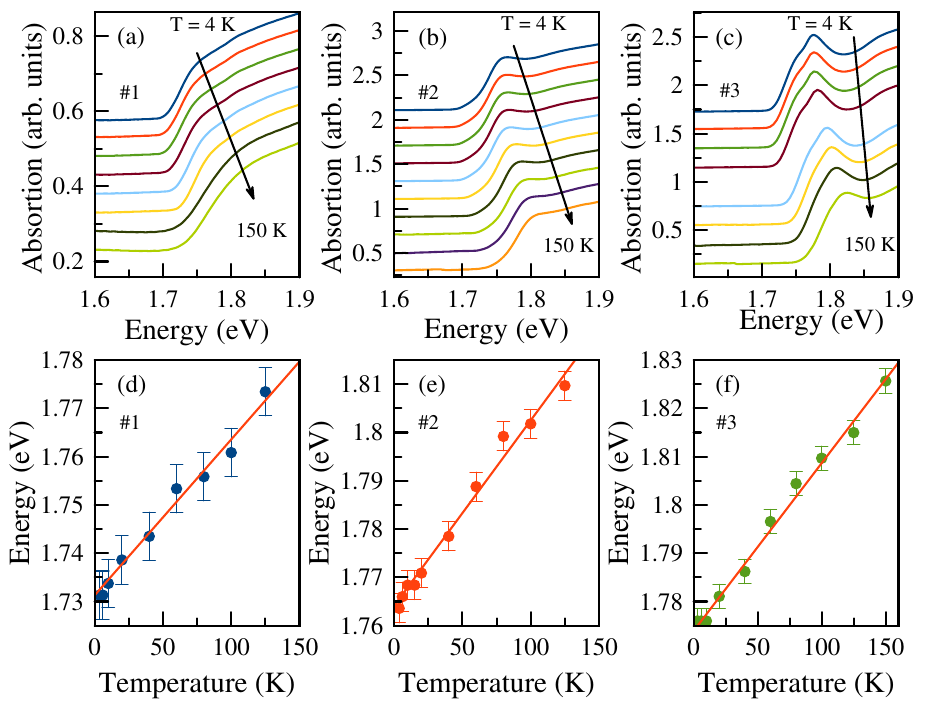}
\caption{Absorption spectra of sample \#1 (a), \#2 (b), and \#3 (c) in the temperature range of $T=4-150$~K. (d),(e),(f) Dependences of the shift of the absorption edge on temperature for samples \#1, \#2 and \#3, respectively. Red lines are linear fits to the measured dependences with parameters given in text.}
\label{figSI:Abs-Temp}
\end{figure}

\subsection{Spin dynamics in different magnetic fields}

Figure~\ref{figSI:MagneticEK31}a plots the FE dynamics in sample \#1 at $T=6$~K for the laser photon energy of 1.722~eV. The spot on the sample on which the measurements are carried out differs from the spot corresponding to Figure~\ref{fig:6K}. Here, the hole spin precession is almost invisible. Therefore, it is possible to evaluate only the electron spin parameters.  Figure~\ref{figSI:MagneticEK31}b shows the magnetic field dependence of the electron Larmor precession frequency. A fit to these data with eq.~\eqref{eq:LarmFreq} gives the electron $g$-factor $g_{\rm e}=2.26$. The magnetic field dependence of the electron spin dephasing time $T_\text{2,e}^*$ is presented in Figure~\ref{figSI:MagneticEK31}c. $T_\text{2,e}^*$ decreases with growing magnetic field from about 800~ps at zero field to 200~ps at the maximum magnetic field of 430~mT. Fitting the data with eq.~\eqref{eq:InhDeph} gives the electron $g$-factor spread  $\Delta g_{\rm e}=0.17$.

\begin{figure}[hbt!]
\includegraphics[width=0.8\columnwidth]{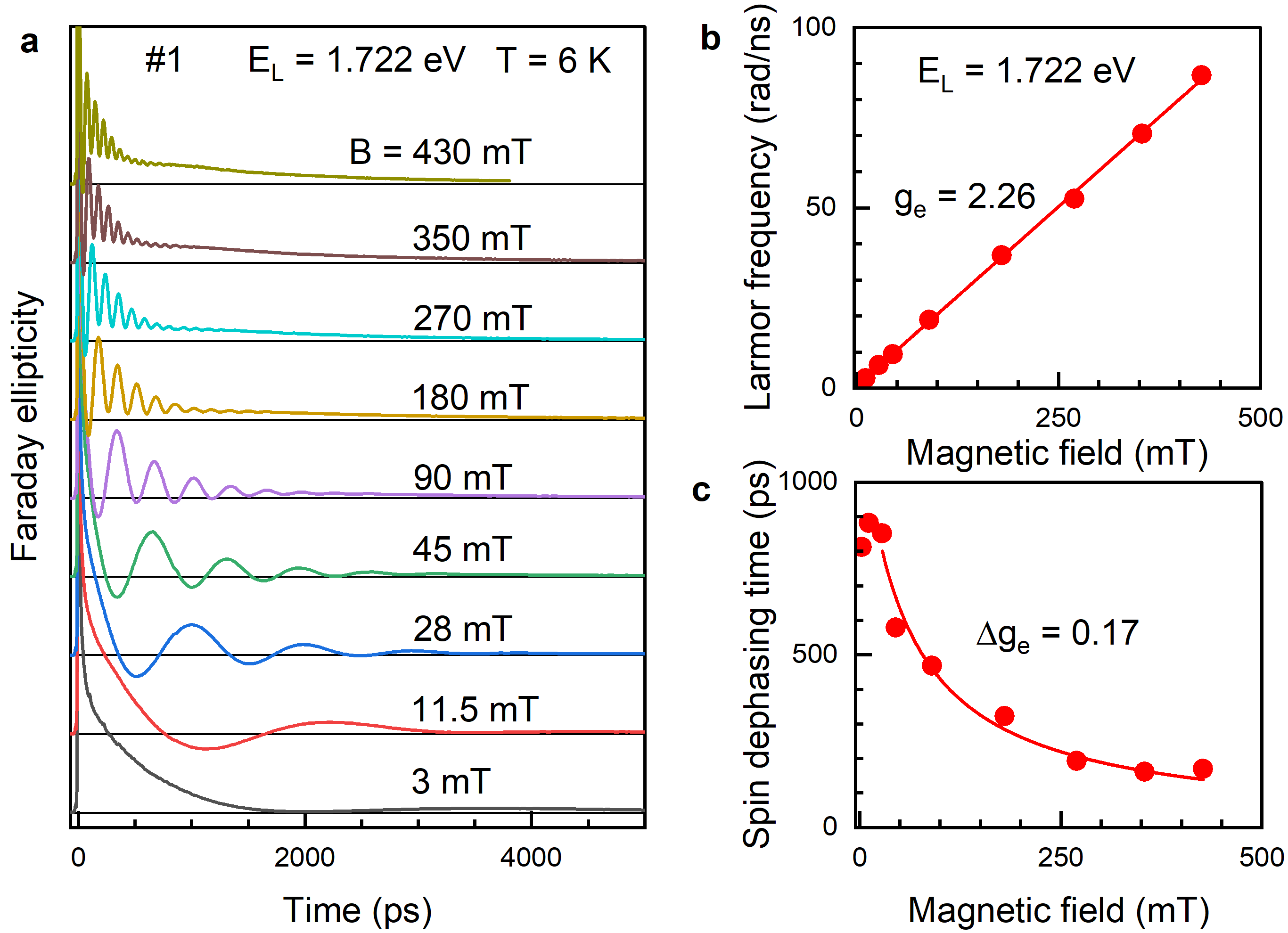}
\caption{Magnetic field dependence of spin dynamics at $T=6$~K for laser photon energy $E_\text{L}=1.722$~eV in sample \#1.
(a) TRFE dynamics measured in various magnetic fields from 3~mT up to 430~mT.
(b) Magnetic field dependence of the electron Larmor precession frequency. Line shows a fit with eq.~\eqref{eq:LarmFreq}. The slope of the fit corresponds to $g_{\rm e}=2.26$.
(c) Magnetic field dependence of the electron spin dephasing time. Fitting the experimental data with eq.~(\ref{eq:InhDeph}) (line) yields $\Delta g_{\rm e}=0.17$.
}
\label{figSI:MagneticEK31}
\end{figure}

\begin{figure}[hbt!]
\includegraphics[width=1\columnwidth]{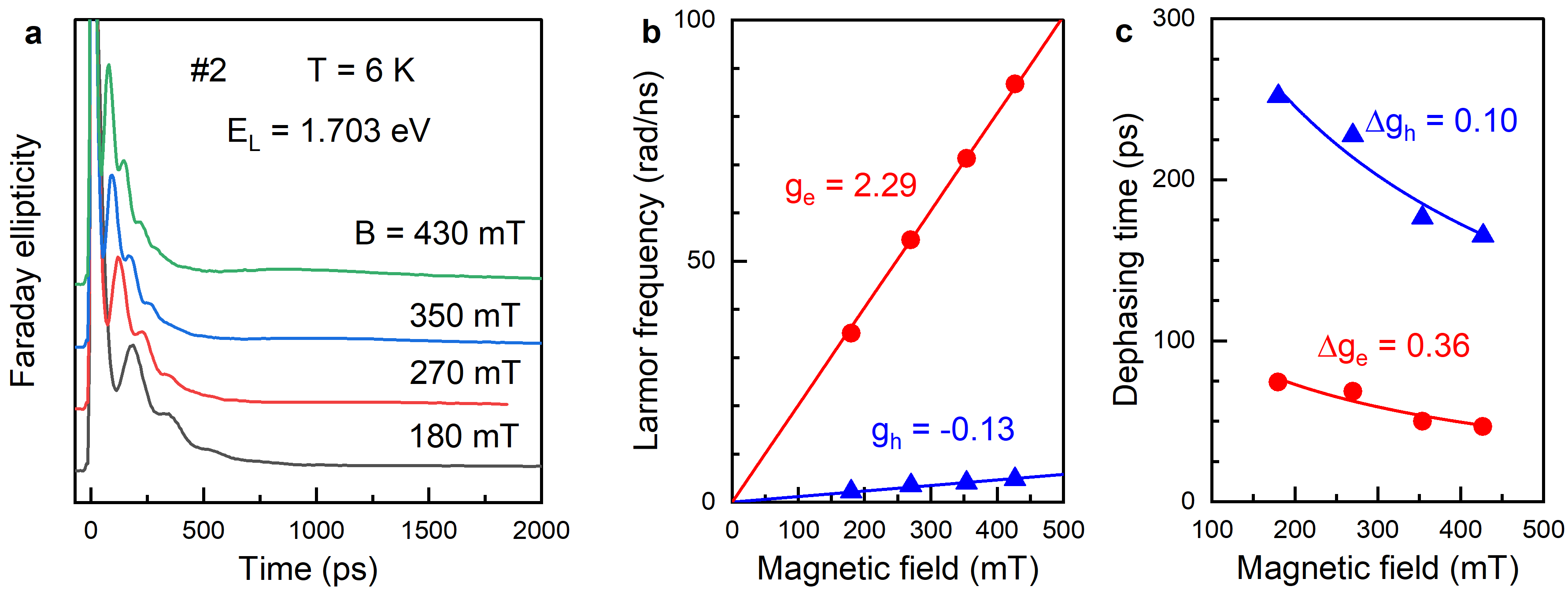}
\caption{Magnetic field dependence of the spin dynamics in sample \#2 measured at $T=6$~K for $E_\text{L}=1.703$~eV.
(a) FE dynamics measured in various magnetic fields from 180~mT up to 430~mT.
(b) Dependences of the hole (blue triangles) and electron (red circles) Larmor precession frequencies on magnetic field. Lines show fits using eq.~\eqref{eq:LarmFreq}. The slopes of the fits correspond to $g_{\rm e}=2.29$ and $g_{\rm h}=-0.13$.
(c) Magnetic field dependences of the electron (red circles) and hole (blue triangles) spin dephasing times. Fits to the experimental data with eq.~\eqref{eq:InhDeph} (lines) yield $\Delta g_{\rm e}=0.36$ and $\Delta g_{\rm h}=0.10$.
}
\label{figSI:EK7}
\end{figure}

Figure~\ref{figSI:EK7} plots similar data for sample \#2, measured at $E_\text{L}=1.703$~eV. Here we observe both electron and hole spin precession. The slopes of the Larmor precession frequency dependences on magnetic field give $g_{\rm e}=2.29$ and $g_{\rm h}=-0.13$ (the sign of the hole $g$-factor is taken according to Figure~\ref{fig:6K}e). The decrease of the electron and hole spin dephasing times with growing magnetic field corresponds to $\Delta g_{\rm e}=0.36$ and $\Delta g_{\rm h}=0.10$, respectively.

\clearpage

\subsection{Sign of the hole $g$-factor}

Figure~\ref{figSI:sign} plots the dependence of the absolute value of the hole $g$-factor on the laser photon energy $E_\text{L}$ (the corresponding spectral dependence of $g_{\rm h}$ is shown in Figure~\ref{fig:6K}e). $|g_{\rm h}|$ behaves in a nonmonotonic way with growing energy: it decreases in the spectral range from 1.69~eV to 1.71~eV, is close to zero at $E_\text{L} \approx 1.71$~eV, and increases in the range from 1.71~eV to 1.78~eV. According to Ref.~[Nestoklon, et al., \textit{Nano Lett.} 2023, \textbf{23}, 8218] $g_{\rm h}$ should monotonically change in this spectral range with increasing confinement energy in CsPbI$_3$ NCs, starting from negative values and rising toward positive values.  Thus, we suggest that $g_{\rm h}<0$ at $E_\text{L}<1.71$~eV, $g_{\rm h}$ crosses zero at $E_\text{L} \approx 1.71$~eV, and $g_{\rm h} >0$ at $E_\text{L}>1.71$~eV. We use this hypothesis for presenting data in the main text, e.g., in Figure~\ref{fig:6K}e.

\begin{figure}[hbt!]
\includegraphics[width=0.5\columnwidth]{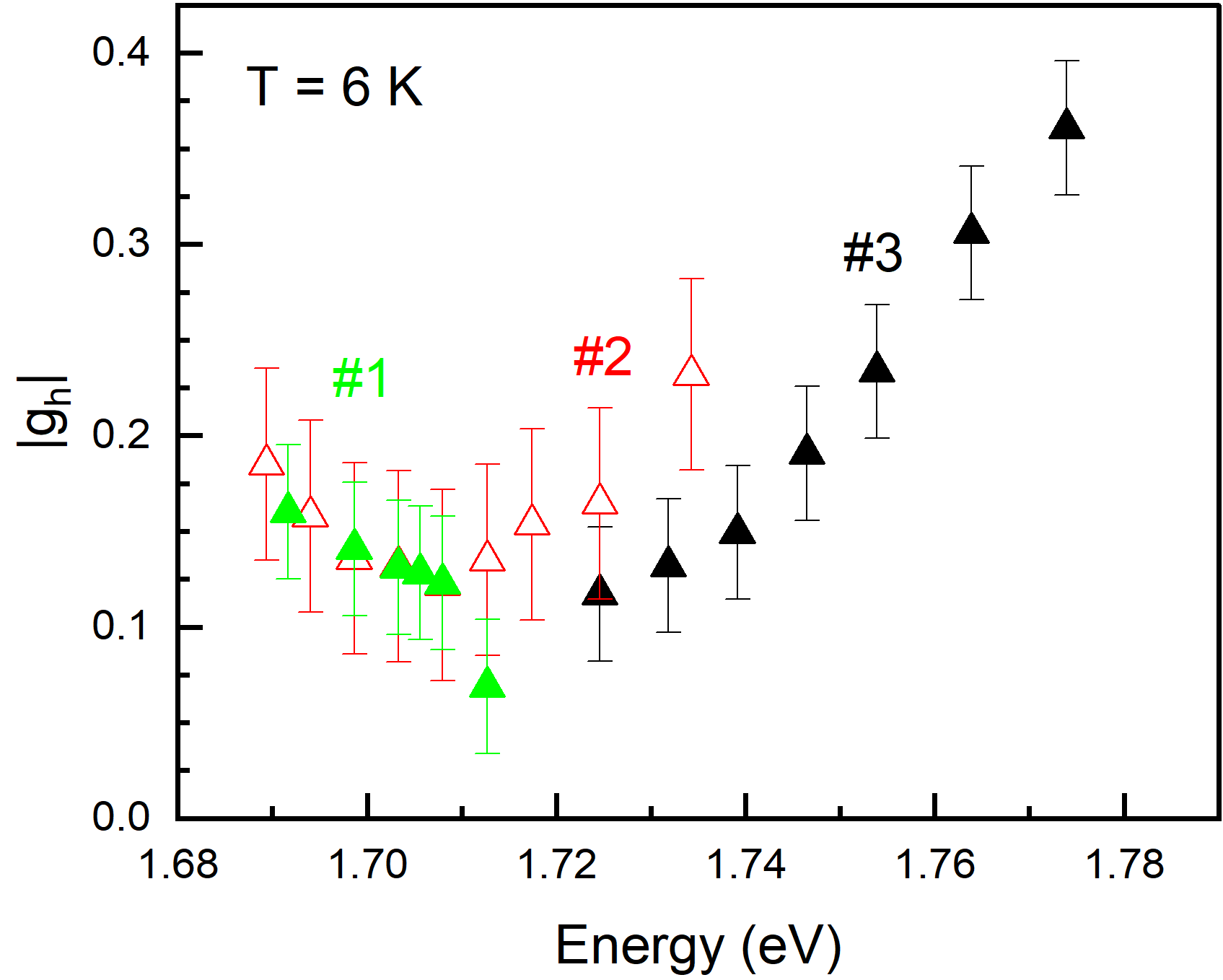}
\caption{Spectral dependence of the absolute value of the hole $g$-factor in samples \#1 (green triangles), \#2 (red open triangles), and \#3 (black triangles). Error bars correspond to the spreads of the $g$-factor distribution $\Delta g_{\rm h} \approx 0.07$ for samples \#1 and \#3  and $\Delta g_{\rm h} \approx 0.10$ for sample \#2.
}
\label{figSI:sign}
\end{figure}

\clearpage

\subsection{Spin dynamics at temperatures up to 120~K}

Figure~\ref{figSI:temp120} shows the temperature dependence of the spin dynamics in sample~\#1. The Voigt magnetic field is equal to 430~mT. These measurements are performed on a sample spot different from that used for the measurements in Figure~\ref{fig:Temperature}. The laser energy $E_\text{L}$ is adjusted to the maximum of the signal for each temperature. Figure~\ref{figSI:temp120}a presents the FE dynamics in sample \#1 at various temperatures from 9~K to 120~K. We do not observe hole spin oscillations up to temperature of about 50~K. Figures~\ref{figSI:temp120}b and~\ref{figSI:temp120}c show the temperature dependences of the electron and hole $g$-factors. Figure~\ref{figSI:temp120}d shows corresponding laser photon energies $E_\text{L}$. The $g$-factor behavior with growing temperature is in agreement with the results presented in the main text.

\begin{figure}[hbt!]
\includegraphics[width=1\columnwidth]{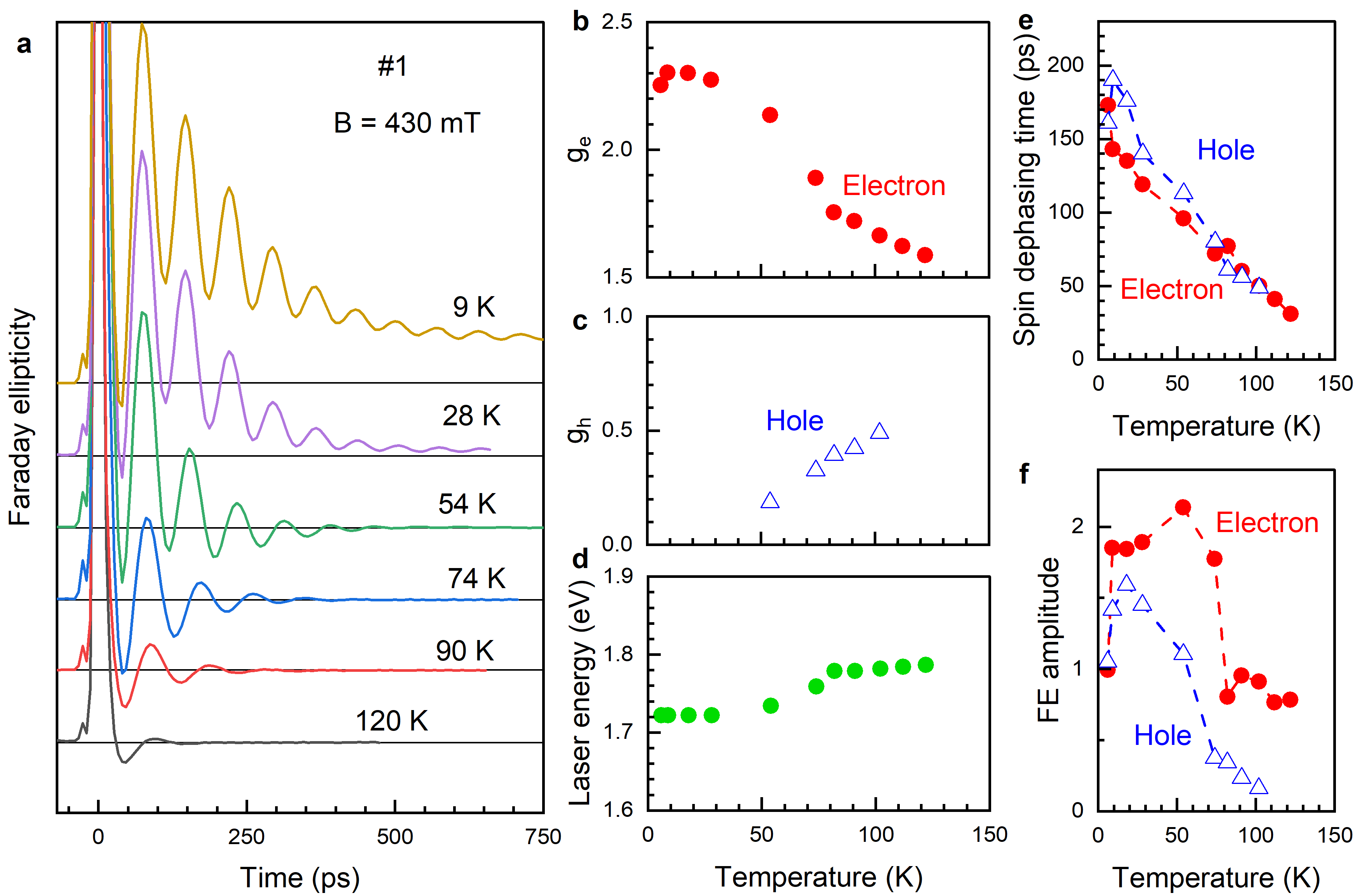}
\caption{Temperature dependence of the spin dynamics in sample \#1 measured in the Voigt magnetic field of 430~mT.
(a) TRFE traces at various temperatures from 9~K to 120~K.  They are shifted vertically for clarity.
(b,c) Temperature dependences of the electron and hole $g$-factors. 
(d) Laser photon energies at which the spin dynamics at various temperatures are measured.
(e) Temperature dependence of the spin dephasing times $T_2^*$ for electrons (red circles) and holes (blue triangles).
(f) Temperature dependence of the FE amplitudes for electrons (red circles) and holes (blue triangles).
}
\label{figSI:temp120}
\end{figure}

\clearpage

\subsection{Theory}\label{sec:SI:Theory}

In this section we estimate the change of the effective band gap, masses, and $g$-factors in CsPbI$_3$ nanocrystals as function of temperature. We start from the bulk material parameters. In Ref.~\cite{Yang2017}, a linear change of the band gap as function of temperature is observed, with the slope for bulk CsPbI$_3$ given by $\dd{E_g}{T} \approx 3.1 \times 10^{-4}$~eVK$^{-1}$. For the analysis, it is more convenient to use this value divided by the band gap
\begin{equation}
  \xi = \frac1{E_g} \dd{E_g}{T} \approx 1.8 \times 10^{-4} \, {\text{K}}^{-1}\,.
\end{equation}
We will also take into account the linear extension of the lattice with temperature
\begin{equation}
  \lambda = \frac1{a_0} \dd{a_0}{T}\,,
\end{equation}
for CsPbI$_3$ $\lambda \approx 3.39 \times 10^{-5}$~K$^{-1}$ \cite{Trots2008}.

As mentioned in Ref.~\cite{Yang2017}, the change of the band gap leads to renormalization of the effective masses of the charge carriers, which may be estimated assuming that this change is small. 
In the {\bf k}$\cdot${\bf p} method, the effective masses of the carriers are \cite{kirstein2022nc}: 
\begin{subequations}\label{eq:theory:masses}
\begin{align}
  \frac{m_0}{m_e} &= 1 + \frac23 \frac{p^2}{m_0 E_g} \,, \\
  \frac{m_0}{m_h} &= -1 + \frac23 \frac{p^2}{m_0 E_g} \frac{3E_g}{E_g+\Delta} \,,
\end{align}
\end{subequations}
where $\Delta$ is the spin-orbit splitting of the conduction band.
Assuming all changes to be small, we may estimate the linear slopes of the mass changes by differentiating Eqs.~\eqref{eq:theory:masses}, giving the result 
\begin{subequations}\label{eq:theory:dmasses}
\begin{align}
  \frac1{m_e}\dd{m_e}{T} &= \left( 1- \frac{m_e}{m_0} \right) \frac1{E_g} \dd{E_g}{T}  \,, \\
  \frac1{m_h}\dd{m_h}{T} &= \left( 1- \frac{m_h}{m_0} \right) \left( 1-\frac{2\Delta}{E_g+\Delta} \right) \frac1{E_g} \dd{E_g}{T} \,. 
\end{align}
\end{subequations}
Using the actual parameters of CsPbI$_3$, where $m_{e,h}\ll m_0$ and $\Delta \approx E_g$, for the estimations it is safe to approximate like
\begin{equation}\label{eq:masses_vs_T}
  \frac1{m_e}\dd{m_e}{T} \approx \frac1{E_g} \dd{E_g}{T}\,,\;\;\;
  \frac1{m_h}\dd{m_h}{T} \approx 0\,.
\end{equation}
Note that in  Ref.~\cite{Yang2017} the authors neglect the contribution to the hole mass from the spin-split electron band and thus severely overestimate its change.


Next, we need to estimate the change of the effective band gap. It may be evaluated as
\begin{equation}
  \frac1{E_{qc}}\dd{E_{qc}}{T} = - \left[ \frac1{m_{e,h}} \dd{m_{e,h}}{T} + 2 \frac{1}{L}\dd{L}{T}  \right]
  = - \left[ \xi + 2 \lambda \right] \,,
\end{equation}
where $L$ is the nanocrystal edge length. Since $\lambda \ll \xi$, the change of the slope of the effective band gap as function of temperature due to confinement is of the order of $E_{qc}/E_g$, which is always small, even for the smallest NCs.
\begin{equation}
  \dd{E_{qc}}{T} \bigg/ \dd{E_g}{T} \sim - \frac{E_{qc}}{E_g} \,.
\end{equation}

Now we proceed to the $g$-factor value change with temperature. In the {\bf k}$\cdot${\bf p} method, its value is given by \cite{nestoklon2023_nl}
\begin{subequations}
\begin{align}
  g_h(E_g,E_h) &= 2-\frac43\frac{p^2w_h }{m_0} \left[ \frac1{E_g+E_h} - \frac1{E_g+E_h+\Delta} \right]\,,\label{eq:theory:gh}\\
  g_e(E_g,E_e) &= -\frac23+\frac43\frac{p^2}{m_0} \frac{w_e}{E_g+E_e} +\Delta g_{\mathrm{rem}} - 40 \frac{m_0}{m_e} \frac{E_e}{\Delta}\,,\label{eq:theory:ge}
\end{align}
\end{subequations}
where $\Delta g_{\mathrm{rem}}$ is the correction from the remote bands and $w_h$, $w_e$ account for the confinement-induced band mixing, see details in \cite{nestoklon2023_nl}. In equations above, in the electron $g$-factor we used a rough estimate for the term arising from the mixing with the spin-orbit split-off electron band, see Supporting Information of Ref.~\cite{nestoklon2023_nl}.

From Eq.~\eqref{eq:theory:gh} it follows that
\begin{equation}
  \dd{g_{h}}{E_g}  = \dd{g_{h}}{E_h}
\end{equation}
and the slope of the hole $g$-factor dependence on energy does not depend on the reason of the energy change, including the change as function of temperature.

For the electron $g$-factor the situation is more complicated. From Eq.~\eqref{eq:theory:ge} it follows that
\begin{subequations}
\begin{align}
  \zeta_e^b \equiv \dd{g_e}{E_g} &= -\frac43\frac{p^2w_e }{m_0} \frac1{(E_g+E_h)^2}\,,\label{eq:theory:dgeEg}\\
  \zeta_e^{qc} \equiv \dd{g_e}{E_e} &= -\frac43\frac{p^2w_e }{m_0} \frac1{(E_g+E_h)^2}-40 \frac{m_0}{m_e} \frac{1}{\Delta}\,.\label{eq:theory:dgeEe}
\end{align}
\end{subequations}
The second term in Eq.~\eqref{eq:theory:dgeEe} is responsible for the deviation of the $g$-factors as function of quantum confinement energy in Fig.~2 of Ref.~\cite{nestoklon2023_nl} from the universal dependence found in Ref.~\cite{kirstein2022nc}. 

Next, we want to understand the evolution of the electron $g$-factor as function of the effective band gap when this change is due to a temperature change. The electron quantum confinement energy may be estimated as 
\begin{equation}
  E_{e} = \frac{\hbar^2}{2m_e} \frac{3\pi^2}{L^2}
\end{equation}
and 
\begin{equation}
  \dd{E_{e}}{T} = - E_e \left[ \xi + 2\lambda \right]\,.
\end{equation}
Let us assume a small change of the temperature ${\rm d}T$, resulting in a change of the electron $g$-factor 
\begin{equation}
  {\rm d} g_e = \left[ \dd{g_e}{E_g} \dd{E_g}{T} + \dd{g_e}{E_e} \dd{E_e}{T} \right] {\rm d}T \,,
\end{equation}
and the peak energy change
\begin{equation}
  {\rm d} E(T) = \dd{E_g}{T}{\rm d}T + \dd{E_{qc}}{T}{\rm d}T  \,.
\end{equation}
The value of interest for us is 
\begin{equation}\label{eq:theory:dgdET}
  \frac{{\rm d} g_e}{{\rm d} E(T)} \approx \zeta_e^b \left(1+\frac{E_{qc}}{E_g}\left(1+2\frac{\lambda}{\xi}\right)\right) 
  - \zeta_e^{qc} \frac{E_e}{E_g} \left(1+2\frac{\lambda}{\xi}\right)\,,
\end{equation}
where we used $E_e,\;E_{qc} \ll E_g$.

Note that for CsPbI$_3$ NCs the ``quantum confinement'' contribution to the $g$-factor change relative to bulk is large: $\zeta_e^{qc} \sim -10$~eV$^{-1}$, while $\zeta_e^{b} \sim -2$~eV$^{-1}$. This means that the renormalization of the electron $g$-factor as a function of energy should be seen already when quantum confinement is small compared to the bulk band gap value. However, the change is expected to be small even for small nanocrystals and the bulk trend should be reproduced, in contrast to the experimental data.

To highlight the difference between the experimental data and the expected trend, we also analyze the change of the $g$-factor as function of the peak energy for different temperatures. For a particular sample, this value is perfectly fitted by a linear function from which $\dd{g_e}{E_g}$ may be extracted from the experimental data, see Fig.~\ref{fig:dgde_vs_T}. As can be seen, this value changes from $-9.2$~eV$^{-1}$ to $-4.2$~eV$^{-1}$ when the temperature changes from 6~K to 120~K. 

\begin{figure}[hbt!]
  \includegraphics[width=0.5\columnwidth]{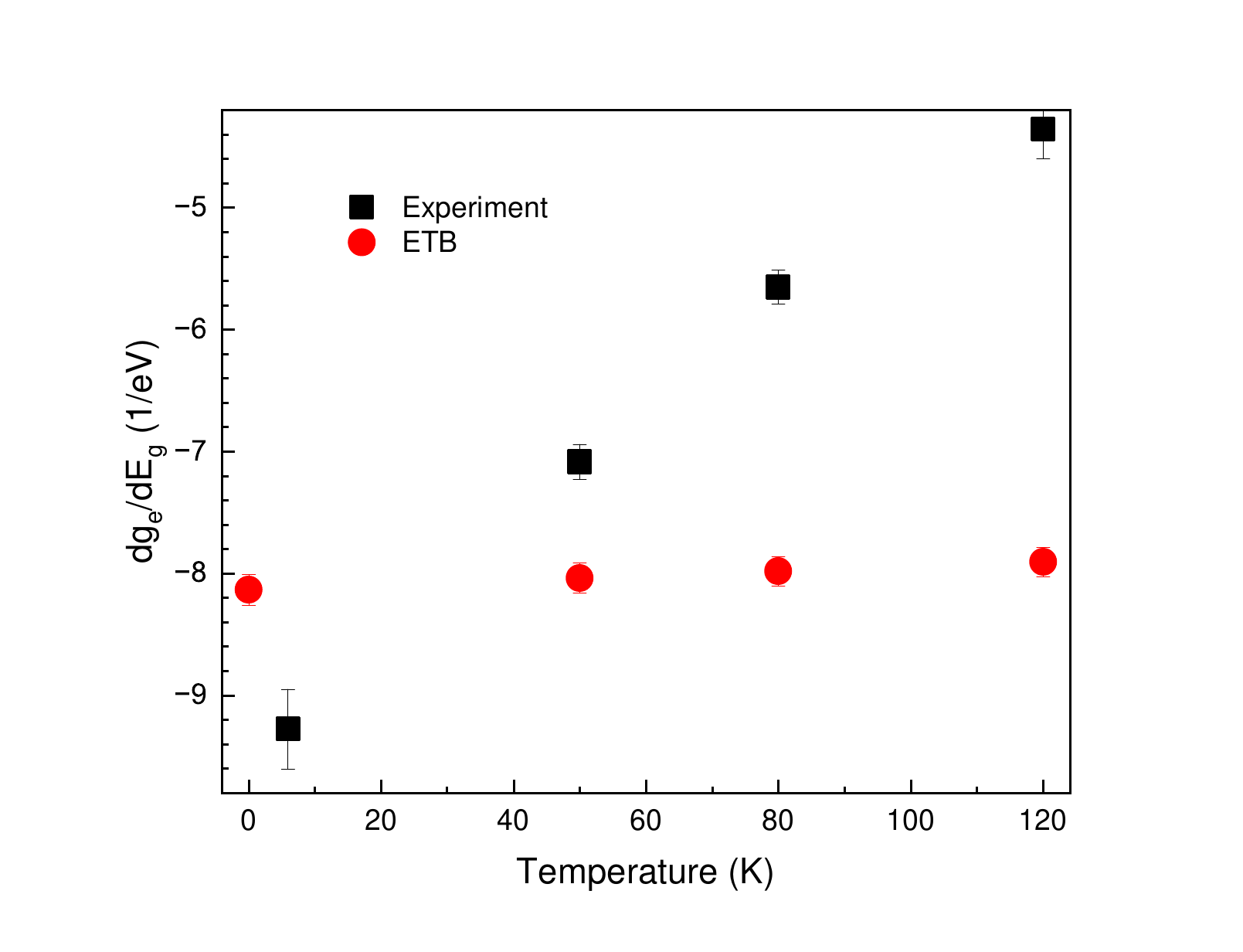}
\caption{Black squares show the experimentally measured $\dd{g_e}{E_g}$ for different temperatures. Red circles show the same value estimated from ETB calculations which take into account only the band gap change and the lattice extension. It is clear that the experimental trend is not reproduced in the atomistic calculations. }
\label{fig:dgde_vs_T}
\end{figure}

To check the theory expectations, we use the empirical tight-binding (ETB) method following Ref.~\onlinecite{nestoklon2023_nl}. To account for the temperature change, we adjust the parameter $E_{pc}$ to reproduce the band gap change. We also change the lattice constant. The bulk values of the effective masses and $g$-factors change qualitatively following Eqs.~\eqref{eq:masses_vs_T} and the results of Ref.~\cite{kirstein2022nc} respectively. However, the value of the $g$-factor derivative with respect to $E_g$ changes only marginally, see Fig.~\ref{fig:dgde_vs_T}. The striking difference between theoretical analysis and experimental results show that an important ingredient is missing in the theoretical analysis. 

\clearpage

\end{widetext}


\section*{References}


\begin{thebibliography}{}



\bibitem{Kovalenko2017}
M. V. Kovalenko, L. Protesescu, and M. I. Bondarchuk,
Properties and potential optoelectronic applications of lead halide perovskite nanocrystals,
\textit{Science} 2017, \textbf{358}, 745--750.	

\bibitem{Efros2021} Al. L. Efros, and L. E. Brus,
Nanocrystal quantum dots: From discovery to modern development,
	\textit{ACS Nano} 2021, \textbf{15}, 6192--6210.

\bibitem{Shamsi2019} J. Shamsi, A. S. Urban, M. Imran, L. De Trizio, and L. Manna,
Metal halide perovskite nanocrystals: synthesis, post-synthesis modifications, and their optical properties,
\textit{Chemical Reviews}, 2019, \textbf{119}, 3296--3348.

\bibitem{Dey2021} A. Dey, J. Ye, A. De, E. Debroye, S. K. Ha, et al.,
State of the art and prospects for halide perovskite nanocrystals,
\textit{ACS Nano}, 2021, \textbf{15}, 10775--10981.

\bibitem{Mu2023} Y. Mu, Z. He, K. Wang, X. Pi, and S. Zhou,
Recent progress and future prospects on halide perovskite nanocrystals for optoelectronics and beyond,
\textit{iScience}, 2023, \textbf{25}, 105371.

\bibitem{Huang2023} C.-Y. Huang, H. Li, Y. Wu, C.-H. Lin, X. Guan, L. Hu, J. Kim, X. Zhu, H. Zeng, and T. Wu,
Inorganic halide perovskite quantum dots: A versatile nanomaterial platform for electronic applications,
\textit{Nano-Micro Lett.}, 2023, \textbf{15}, 16.



\bibitem{Li2017}
P. Li, C. Hu, L. Zhou, J. Jiang, Y. Cheng, M. He, X. Liang, and W. Xiang,
Novel synthesis and optical characterization of CsPb$_2$Br$_5$ quantum dots in borosilicate glasses,
\textit{Mater. Lett.}, 2017, \textbf{209}, 483--485.

\bibitem{Liu2018}
S. Liu, Y. Luo, M. He, X. Liang, and W. Xiang,
Novel CsPbI$_3$ QDs glass with chemical stability and optical properties,
\textit{J. Eur. Ceram. Soc.}, 2018, \textbf{38}, 1998--2004.

\bibitem{Liu2018a}
S. Liu, M. He, X. Di, P. Li, W. Xiang, and X. Liang,
Precipitation and tunable emission of cesium lead halide perovskites (CsPbX$_3$, X = Br, I) QDs in borosilicate glass,
\textit{Ceram. Int.}, 2018, \textbf{44}, 4496--4499.

\bibitem{Ye2019}
Y. Ye, W. Zhang, Z. Zhao, J. Wang, C. Liu, Z. Deng, X. Zhao, and J. Han,
Highly luminescent cesium lead halide perovskite nanocrystals stabilized in glasses for light-emitting applications,
\textit{Adv. Opt. Mater.}, 2019, \textbf{7}, 1801663.

\bibitem{Kolobkova2021}
E. V. Kolobkova, M. S. Kuznetsova,  and N. V. Nikonorov,
Perovskite CsPbX$_3$ (X = Cl, Br, I) nanocrystals in fluorophosphate glasses,
\textit{J. Non-Crystalline Solids}, 2021, \textbf{563}, 120811.

\bibitem{Belykh2022} V. V. Belykh,  M. L.  Skorikov, E. V. Kulebyakina,  E. V. Kolobkova, M. S.  Kuznetsova,  M. M. Glazov, and D. R. Yakovlev,
Submillisecond spin relaxation in CsPb(Cl,Br)$_3$ perovskite nanocrystals in a glass matrix,
\textit{Nano Lett.}  2022, \textbf{22}, 4583--4588.



\bibitem{Vardeny2022_book} \textit{Hybrid Organic Inorganic Perovskites: Physical Properties and Applications}. (eds. Z.~V. Vardeny,  M.~C. Beard) (World Scientific, 2022).

\bibitem{Nestoklon2018}
M. O. Nestoklon,  S. V.  Goupalov, R. I.  Dzhioev,  O. S.  Ken, V. L. Korenev,  Yu. G.  Kusrayev, V. F.   Sapega,  C.  de Weerd,  L.  Gomez,  T.  Gregorkiewicz,  J.  Lin,  K.  Suenaga,  Y.  Fujiwara,  L. B.  Matyushkin, and I. N. Yassievich,
 Optical orientation and alignment of excitons in ensembles of inorganic perovskite nanocrystals,
\textit{Phys. Rev. B}  2018, \textbf{97}, 235304.

\bibitem{canneson2017}
D.~Canneson, E.~V. Shornikova, D.~R. Yakovlev, T.~Rogge, A.~A. Mitioglu, M.~V. Ballottin, P.~C.~M. Christianen, E.~Lhuillier, M.~Bayer, and L.~Biadala,
Negatively charged and dark excitons in CsPbBr$_3$ perovskite nanocrystals revealed by high magnetic fields,
\textit{Nano Lett.}  2017, \textbf{17}, 6177--6183.


\bibitem{Crane2020} M. J. Crane,  L. M.  Jacoby,  T. A.  Cohen,  Y. Huang,  C. K. Luscombe, and D. R. Gamelin,
Coherent spin precession and lifetime-limited spin dephasing in CsPbBr$_3$ perovskite nanocrystals,
\textit{Nano Lett.}  {2020}, \textbf{20}, 8626--8633.

\bibitem{Grigoryev2021} P. S. Grigoryev,  V. V.  Belykh,  D. R.  Yakovlev,  E. Lhuillier, and M. Bayer,
Coherent spin dynamics of electrons and holes in CsPbBr$_3$ colloidal nanocrystals,
\textit{Nano Lett.}  {2021}, \textbf{21}, 8481--8487.

\bibitem{Lin2022} X. Lin,  Y.  Han, J.  Zhu, and K. Wu,
Room-temperature coherent optical manipulation of hole spins in solution-grown perovskite quantum dots,
\textit{Nat. Nanotechnol.}  {2022}, \textbf{18}, 124.

\bibitem{Meliakov2023NCs} S. R. Meliakov, E. A. Zhukov, E. V. Kulebyakina,  V. V. Belykh, and D. R. Yakovlev,
Coherent spin dynamics of electrons in CsPbBr$_3$ perovskite nanocrystals at room temperature,
\textit{Nanomaterials}, 2023, \textbf{13}, 2454.

\bibitem{kirstein2023_SML}
E. Kirstein, N.~E. Kopteva, D.~R. Yakovlev, E.~A. Zhukov, E.~V. Kolobkova, M.~S. Kuznetsova, V.~V. Belykh, I.~A. Yugova, M.~M. Glazov, M.~Bayer, and A. Greilich,
Mode locking of hole spin coherences in CsPb(Cl,Br)$_3$ perovskite nanocrystals,
\textit{Nat. Commun.}  {2023}, \textbf{14}, 699.

\bibitem{Strohmair2020} S. Strohmair,  A. Dey, Y. Tong, L. Polavarapu, B. J.  Bohn, and J. Feldmann,
Spin polarization dynamics of free charge carriers in CsPbI$_3$ nanocrystals,
\textit{Nano Lett.}  {2020},  \textbf{20}, 4724--4730.

\bibitem{kirstein2022nc}
E.~Kirstein, D.~R.~Yakovlev, M.~M.~Glazov, E.~A.~Zhukov, D.~Kudlacik, I.~V.~Kalitukha, V.~F.~Sapega, G.~S.~Dimitriev, M.~A.~Semina, M.~O.~Nestoklon, E.~L.~Ivchenko, N.~E.~Kopteva, D.~N.~Dirin, O.~Nazarenko, M.~V.~Kovalenko, A.~Baumann, J.~H\"ocker, V.~Dyakonov, and M.~Bayer,
The Land\'e factors of electrons and holes in lead halide perovskites: universal dependence on the band gap,
\textit{Nat. Commun.} {2022}, \textbf{13}, {3062}.

\bibitem{Kopteva_gX_2023} N. E. Kopteva,  D. R.  Yakovlev, E.  Kirstein,  E. A. Zhukov, D.  Kudlacik,   I. V. Kalitukha,  V. F. Sapega, D. N.  Dirin,  M. V. Kovalenko, A.  Baumann,  J. H\"ocker, V.  Dyakonov,  S. A. Crooker, and M. Bayer,
Weak dispersion of exciton Land\'e  factor with band gap energy in lead halide perovskites: Approximate compensation of the electron and hole dependences,
\textit{Small} 2023, 2300935.

\bibitem{nestoklon2023_nl}
M. O. Nestoklon, E. Kirstein, D. R. Yakovlev, E. A. Zhukov, M. M. Glazov, M. A. Semina, E. L. Ivchenko, E. V. Kolobkova, M. S. Kuznetsova, and M. Bayer,
Tailoring the electron and hole Land\'e factors in lead halide perovskite nanocrystals by quantum confinement and halide exchange,
\textit{Nano Lett.} 2023, \textbf{23}, 8218--8224.


\bibitem{oestreich1995}  M. Oestreich, and W. W. R\"uhle,
Temperature dependence of the electron Land\'e g-factor in GaAs,
\textit{Phys. Rev. Lett.}  {1995}, \textbf{74}, 2315.

\bibitem{oestreich1996}  M. Oestreich,  S. Hallstein,  A. P. Heberle,  K. Eberl, E.  Bauser, and W. W. R\"uhle,
Temperature and density dependence of the electron Land\'e g factor in semiconductors,
\textit{Phys. Rev. B}  {1996}, \textbf{53}, 7911.

\bibitem{Zawadzki2008}  W. Zawadzki,  P. Pfeffer,  R. Bratschitsch,  Z. Chen,  S. T. Cundiff, B. N.  Murdin,  and  C. R. Pidgeon,
Temperature and density dependence of the electron Land\'e g factor in semiconductors,
\textit{Phys. Rev. B}  {2008}, \textbf{78}, 245203.

\bibitem{Hubner2009}  J. H\"ubner,  S. D\"ohrmann, D.  H\"agele, and M. Oestreich,
Temperature-dependent electron Land\'e g factor and the interband matrix element of GaAs,
\textit{Phys. Rev. B}  {2009}, \textbf{79}, 193307.

\bibitem{Yakovlev_Ch6}
D.~R. Yakovlev and M. Bayer,
Coherent spin dynamics of carriers. In \textit {Spin Physics in Semiconductors}, M.~I. Dyakonov (ed.) (Springer International Publishing AG, 2017) chapter 6, pp. 155--206.

\bibitem{belykh2019}
V. V. Belykh, D. R. Yakovlev, M. M. Glazov, P. S. Grigoryev, M. Hussain, J. Rautert, D. N. Dirin, M. V. Kovalenko, and M. Bayer,
Coherent spin dynamics of electrons and holes in CsPbBr$_3$ perovskite crystals,
\textit{Nat. Commun.} {2019}, \textbf{10}, 673.

\bibitem{YugovaPRB09} I. A. Yugova,  M. M. Glazov,  E. L. Ivchenko, and  Al. L. Efros,
Pump-probe Faraday rotation and ellipticity in an ensemble of singly charged quantum dots,
\textit{Phys. Rev. B} {2009}, \textbf{80}, {104436}.

\bibitem{Glazov2010} M. M. Glazov, I. A. Yugova, S. Spatzek, A. Schwan, S. Varwig, D. R. Yakovlev, D. Reuter, A. D. Wieck, and M. Bayer,
Effect of pump-probe detuning on the Faraday rotation and ellipticity signals of mode-locked spins in (In,Ga)As/GaAs quantum dots,
\textit{Phys. Rev. B} 2010, \textbf{82}, 155325.

\bibitem{kirstein2022am}
E. Kirstein, D. R. Yakovlev, M. M. Glazov, E. Evers, E. A. Zhukov, V. V. Belykh, N. E. Kopteva, D. Kudlacik, O. Nazarenko, D. N. Dirin, M. V. Kovalenko, and M. Bayer,
Lead-dominated hyperfine interaction impacting the carrier spin dynamics in halide perovskites,
\textit{Advanced Materials} {2022}, \textbf{34}, 2105263.


\bibitem{Yang2017}
  Z. Yang, A. Surrente, K. Galkowski, A. Miyata, O. Portugall, R. J. Sutton, A. A. Haghighirad, H. J. Snaith, D. K. Maude, P. Plochocka, and R. J. Nicholas,
  Impact of the halide cage on the electronic properties of fully inorganic cesium lead halide perovskites,
  \textit{ACS Energy Lett.} {2017}, \textbf{2}, 1621.

\bibitem{Trots2008}
  D. M. Trots and S. V. Myagkota,
  High-temperature structural evolution of caesium and rubidium triiodoplumbates,
  \textit{J. Phys. Chem. Sol.} {2008}, \textbf{69}, 2520.



\end{thebibliography}
\end{document}